\pdfoutput=1
\documentclass[fleqn,usenatbib,usedcolumn]{mnras}

\usepackage{lscape}
\usepackage{rotating}
\usepackage{ifpdf}

\usepackage{caption}
\usepackage[T1]{fontenc}

\DeclareCaptionFormat{cont}{#1 (cont.)#2#3\par}

\usepackage[T1]{fontenc}
\usepackage{ae,aecompl}

\usepackage{graphicx}	
\usepackage{amsmath}	
\usepackage{amssymb}	
\usepackage{epstopdf}
\usepackage{fixltx2e}
\MakeRobust{\overrightarrow}

\title[Atomic Gas in Debris Discs]{Atomic Gas in Debris Discs}

\author[Hales et al.]{
Antonio S. Hales,$^{1,2}$\thanks{E-mail: ahales@alma.cl}
M.  J. Barlow,$^{3}$, I. A. Crawford$^{4}$ and S. Casassus$^{5,6}$
\\
$^{1}$Atacama Large Millimeter/Submillimeter Array, Joint ALMA Observatory, Alonso de C\'ordova 3107, Vitacura 763-0355, Santiago - Chile\\
$^{2}$National Radio Astronomy Observatory, 520 Edgemont Road, Charlottesville, Virginia, 22903-2475, United States\\
$^{3}$Department of Physics and Astronomy, University College London, Gower Street, London WC1E 6BT, UK\\
$^{4}$Department of Earth and Planetary Sciences, Birkbeck College London, Malet Street, London, WC1E 7HX, UK\\
$^{5}$Departamento de Astronom\'ia, Universidad de Chile, Camino El Observatorio 1515, Las Condes, Santiago, Chile\\
$^{6}$Millennium Nucleus `Protoplanetary Discs', Santiago, Chile 0000-0002-0433-9840\\
}

\date{Accepted 2016 December 13. Received 2016 December 2; in original form 2016 November 22.}

\pubyear{2016}

\begin{document}
\label{firstpage}
\pagerange{\pageref{firstpage}--\pageref{lastpage}}
\maketitle

\begin{abstract}

We have conducted a search for optical circumstellar absorption lines
in the spectra of 16 debris disc host stars. None of the stars in our
sample showed signs of emission line activity in either H$\alpha$,
Ca~{\sc ii} or Na~{\sc i}, confirming their more evolved nature. Four
stars were found to exhibit narrow absorption features near the cores
of the photospheric Ca~{\sc ii} and Na~{\sc i} D lines (when Na~{\sc
  i} D data were available).  We analyse the characteristics of these
spectral features to determine whether they are of circumstellar or
interstellar origins. The strongest evidence for circumstellar gas is
seen in the spectrum of HD~110058, which is known to host a debris
disc observed close to edge-on. This is consistent with a recent ALMA
detection of molecular gas in this debris disc, which shows many
similarities to the $\beta$ Pictoris system. \\

\end{abstract}




\section{Introduction}

Debris discs represent the final stage of the planet formation
process. The dust observed in these systems is thought to be
replenished through collisions between solid bodies \citep{wyatt2008},
and are therefore expected to be gas depleted. However, a few debris
discs show the presence of small amounts of circumstellar gas. Whether
this gas is remnant of the early disc stages (primordial) or of
secondary origin (e.g. brought to the gas phase by photodesorption or
by cometary activity) is a matter of contemporary debate
\citep{Dent14,moor2015,kospal2016,kral2016,marino2016,greaves2016,Lieman2016}.

Circumstellar gas was first discovered around $\beta$~Pictoris over 40
years ago when \citet{Slettebak1975} noticed the presence of narrow
Ca~{\sc ii} absorption lines located at the centre of $\beta$~Pic
photospheric lines. Subsequent to the first imaging of the debris disc
around $\beta$~Pic by \citet{smith84}, \cite{hobbs85} observed similar
features both in Ca~{\sc ii} and Na~{\sc i}. The radial velocities of
the absorptions coincided with the radial velocity of the star, and
were attributed to circumstellar gas orbiting the star. Variable
absorption features have been  indentified, attributed to the
evaporation of solid kilometer-sized bodies falling into the
star \citep{beust90,crawford94,vidal98,kiefer2014a,welsh2016}.

Over a dozen other debris discs with circumstellar gas absorption
features have been indentified with this method, such as HR~10
\citep{lagrange90}, HD~32297 \citep{redfield07b}, 49~Ceti
\citep{Montgomery2012}, HD~172555 \citep{kiefer2014b}, HD~21620,
HD~110411, HD~145964, HD~183324 \citep{welsh2013}, and $\phi$~Leo
\citep{eiroa2016}. Detecting gas in discs using absorption lines is
difficult since the gas disc must be favorably oriented close to
edge-on in order to be detected. On the other hand, the advantage of
this method is that it is sensitive to much lower gas column densities
compared to direct observations of gas emission.

In this work we study the optical spectra of 16 debris disc host stars
in order to search for gas-bearing $\beta$ Pictoris-like systems.  One
of the main difficulties of this technique is to rule out interstellar
(IS) absorption features, which may be very similar in profiles and
strength to circumstellar lines. This requires careful analysis of the
line characteristics
\citet[][]{lagrange90b,crawford98,redfield07b,welsh2013}.

\section{Target sample}

The sample consists of 16 debris disc host stars taken from the
catalogue of \citet{mann98} having spectral types ranging from B8 to
F7 \citep[as listed in the Michigan Spectral Catalog for southern
  stars; ][]{houk88}. The sample was selected in terms of their low
dust fractional luminosities, with all stars having L$_{\rm
  IR}$/L$_{\rm \star}$ values less than those investigated by
\cite{dun97b} being selected in order to choose true debris discs.
Dust fractional luminosities were computed by fitting a black-body to
the observed spectral energy distributions (SED). The stellar SEDs
were approximated using templates from the library of stellar
atmospheres by \citet{castelli2004}, for models with $log g =4.0$ and
solar metallicity.

\begin{figure*}
  \centering
 \includegraphics[scale=0.28,angle=0,clip=true]{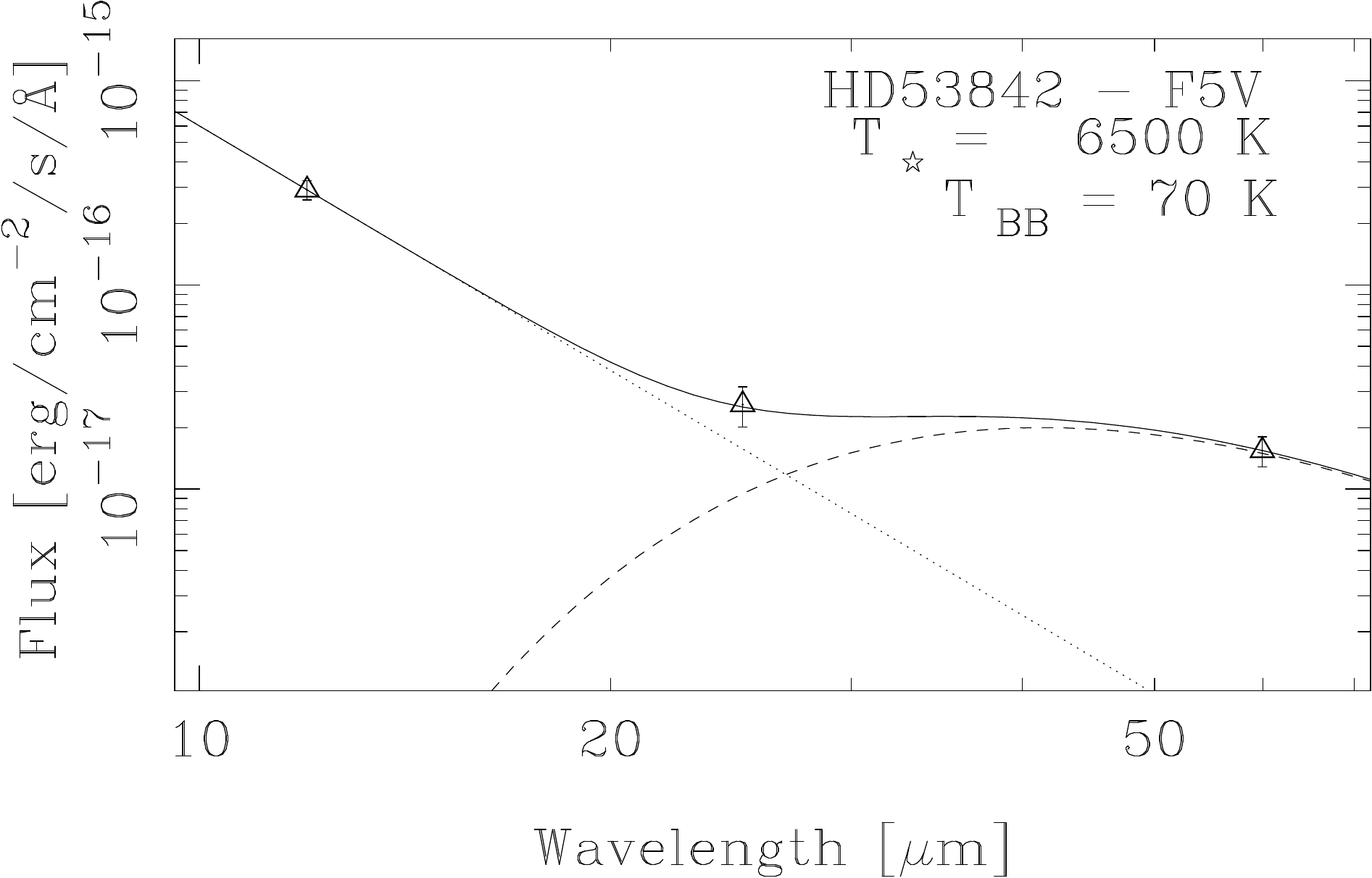}
 \includegraphics[scale=0.28,angle=0,clip=true]{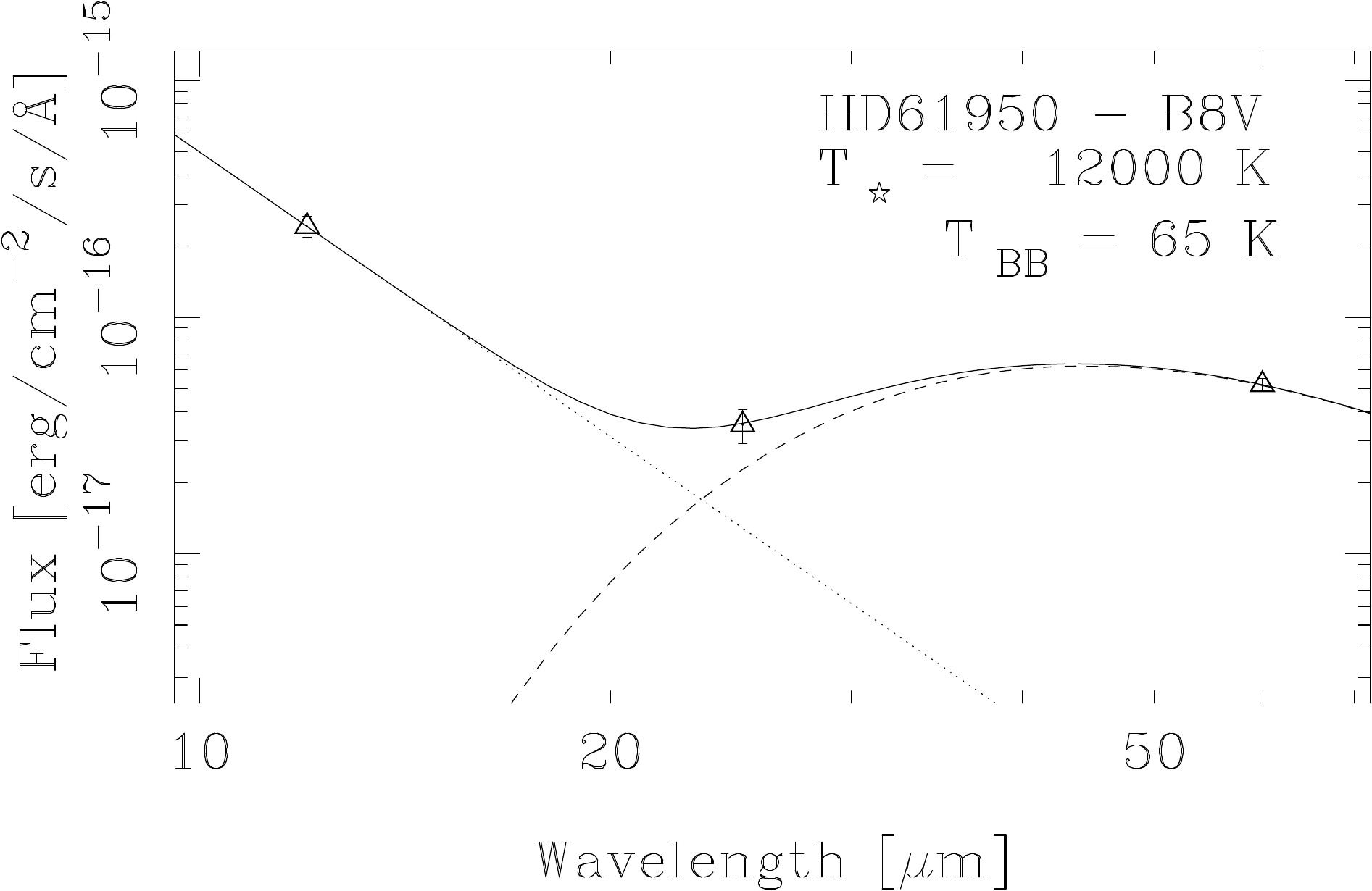}
 \includegraphics[scale=0.28,angle=0,clip=true]{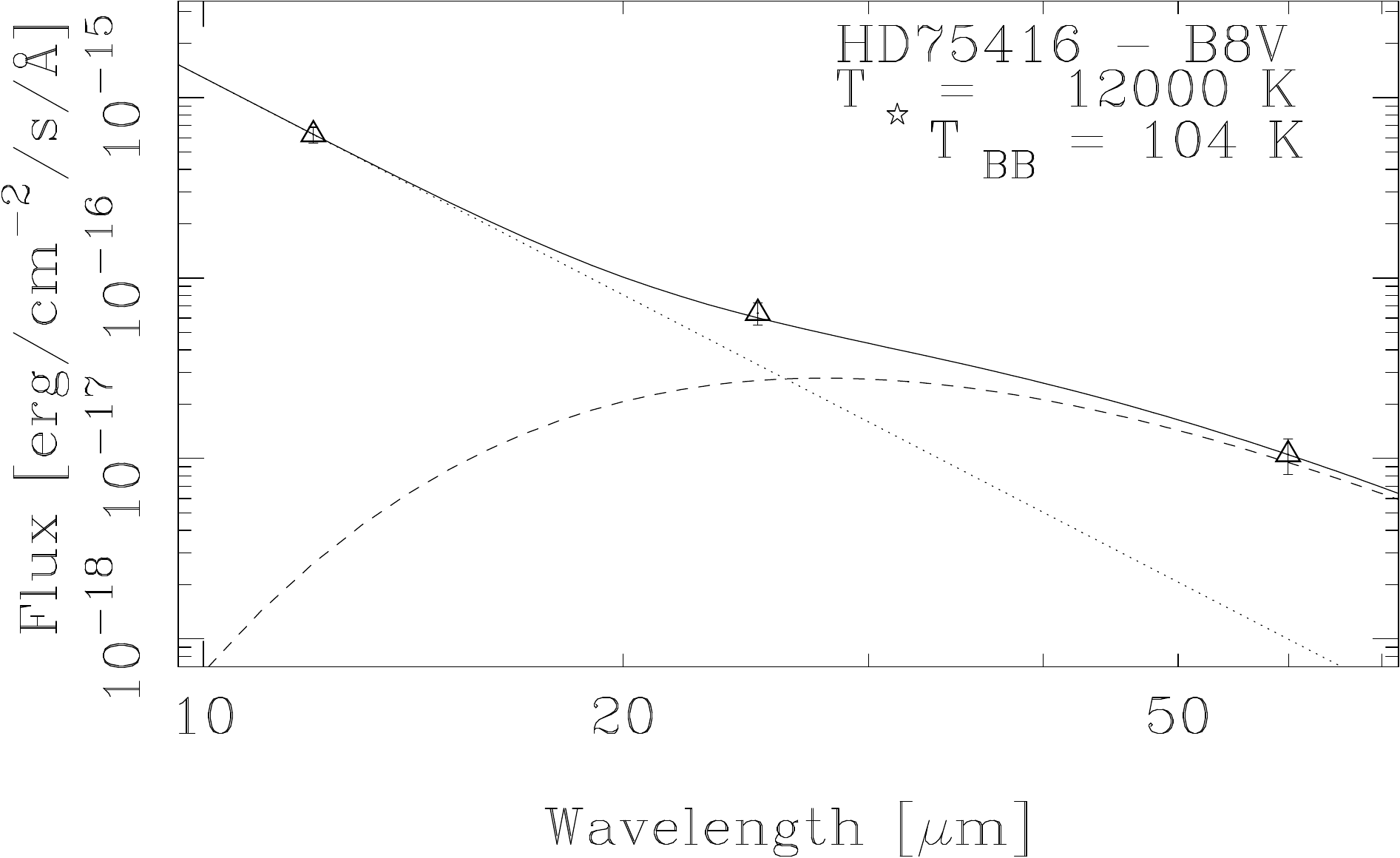}
 \includegraphics[scale=0.28,angle=0,clip=true]{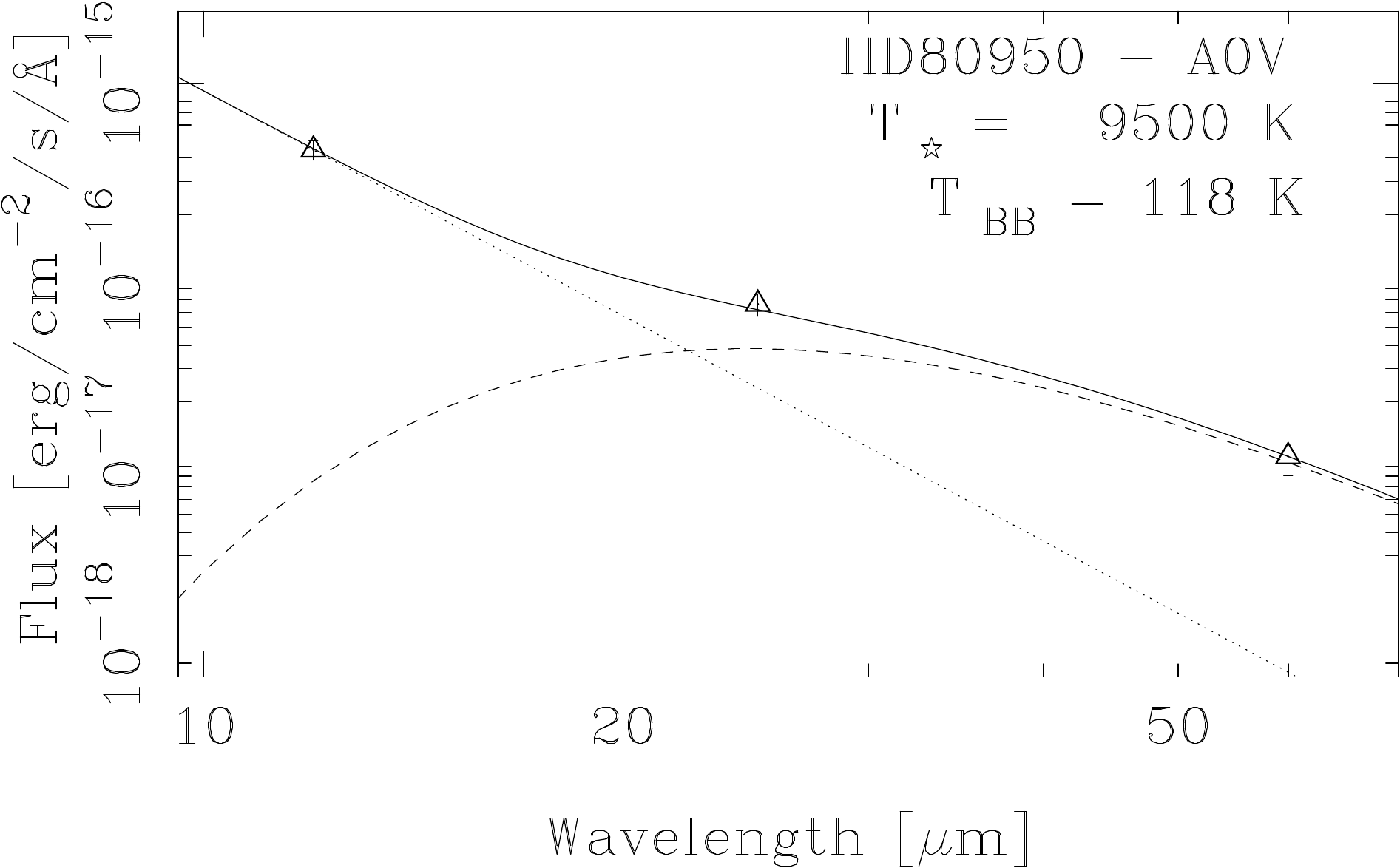}
 \includegraphics[scale=0.28,angle=0,clip=true]{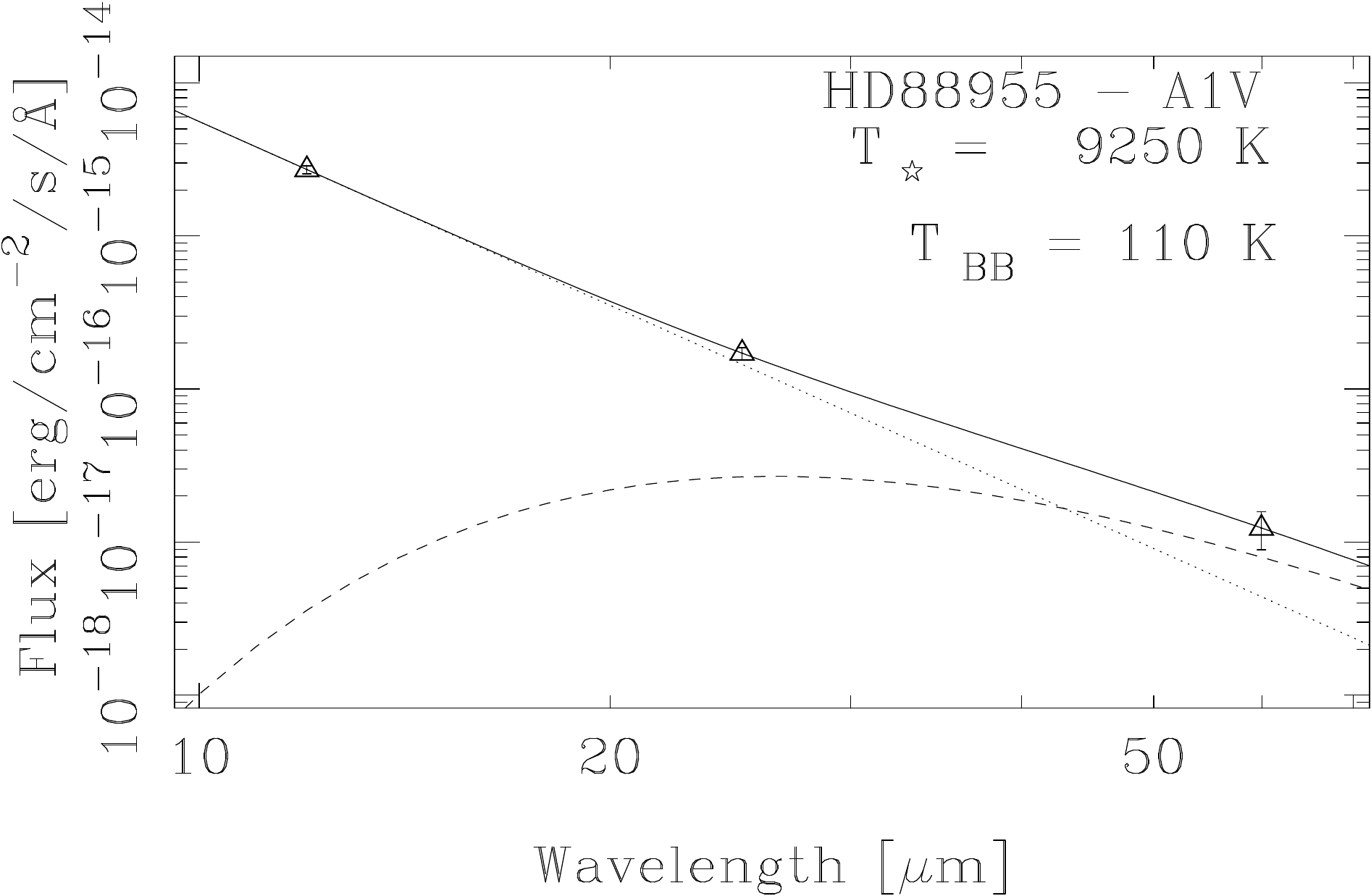}
 \includegraphics[scale=0.28,angle=0,clip=true]{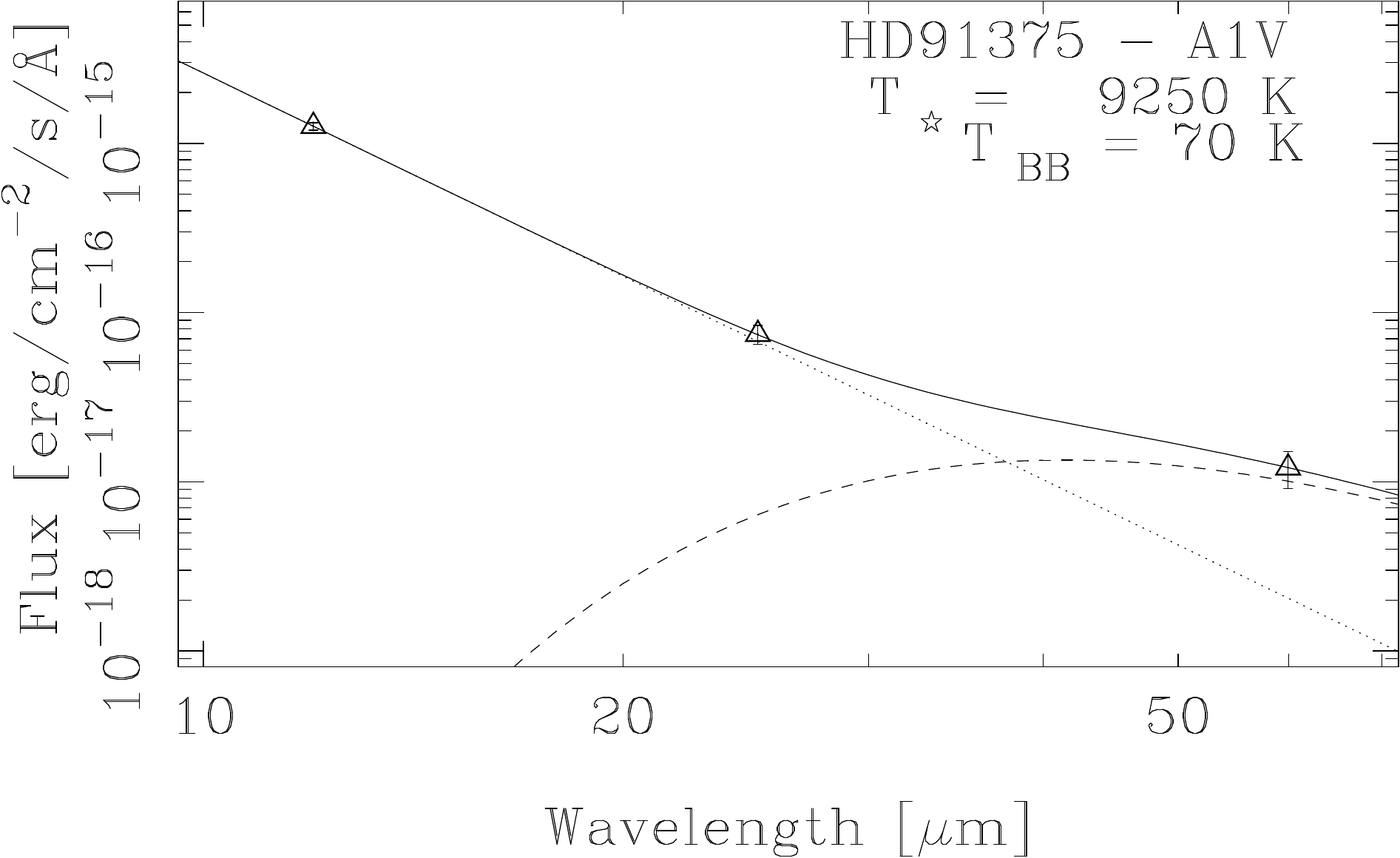}
 \includegraphics[scale=0.28,angle=0,clip=true]{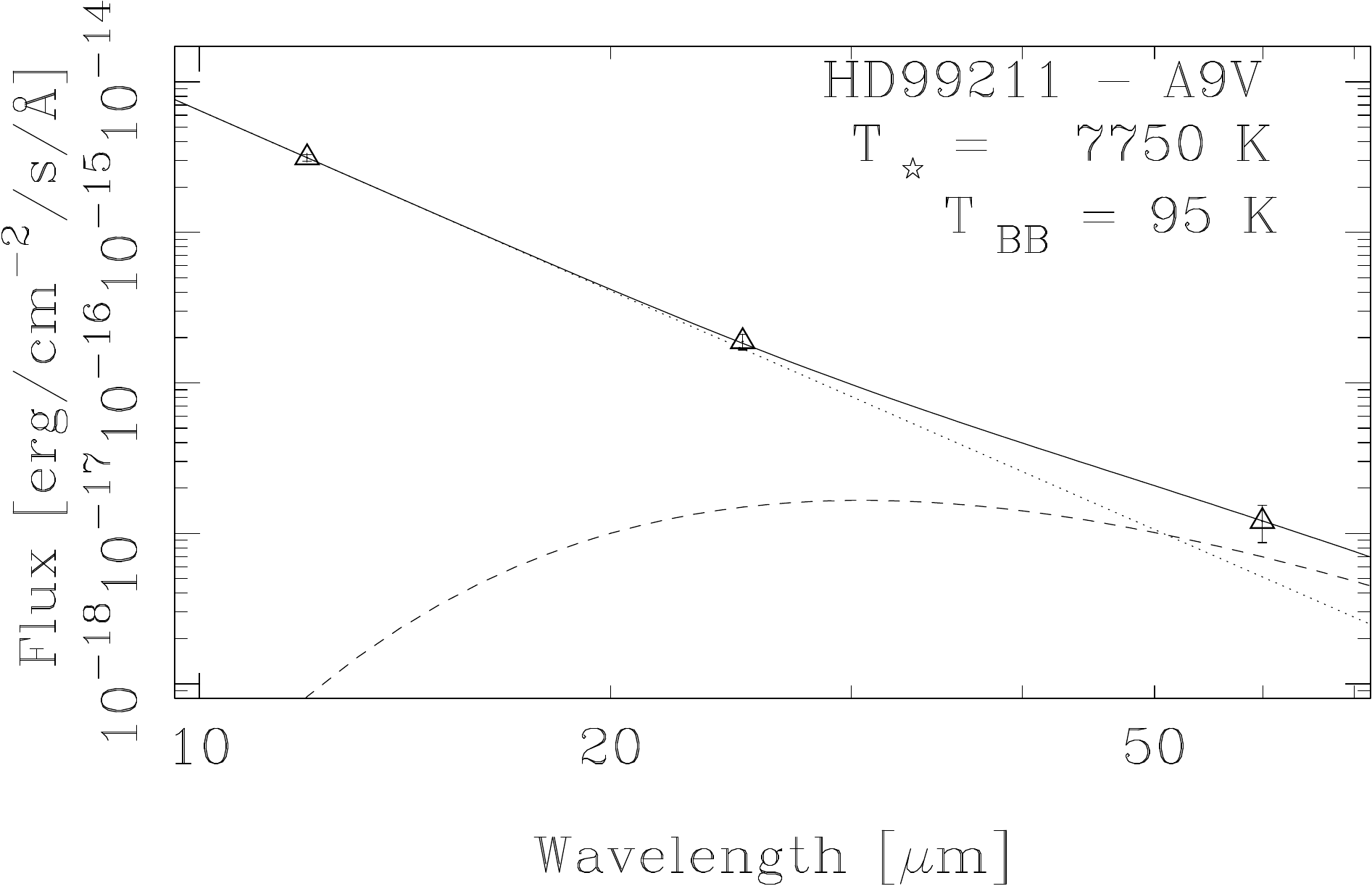}
 \includegraphics[scale=0.28,angle=0,clip=true]{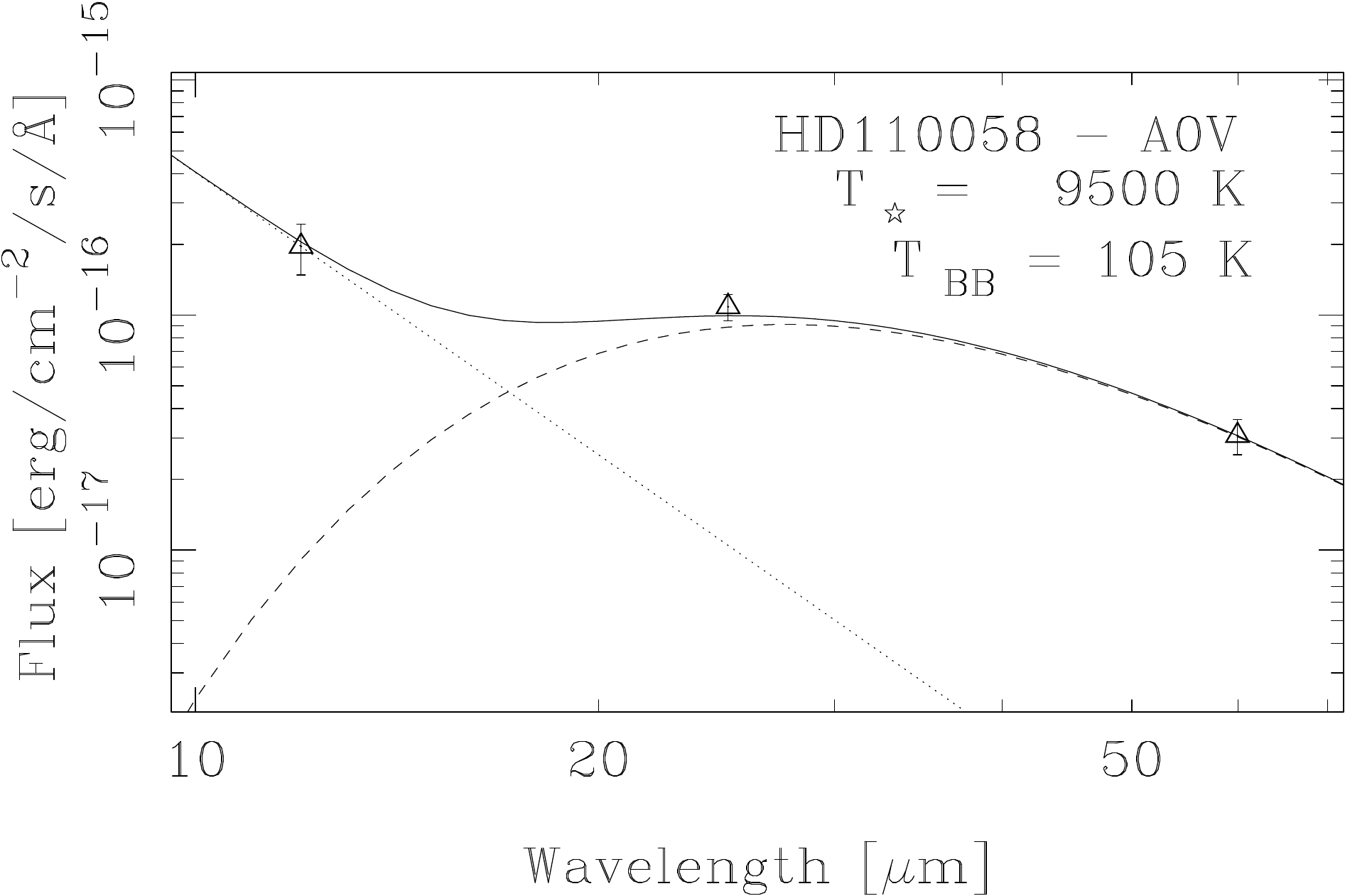}
\includegraphics[scale=0.28,angle=0,clip=true] {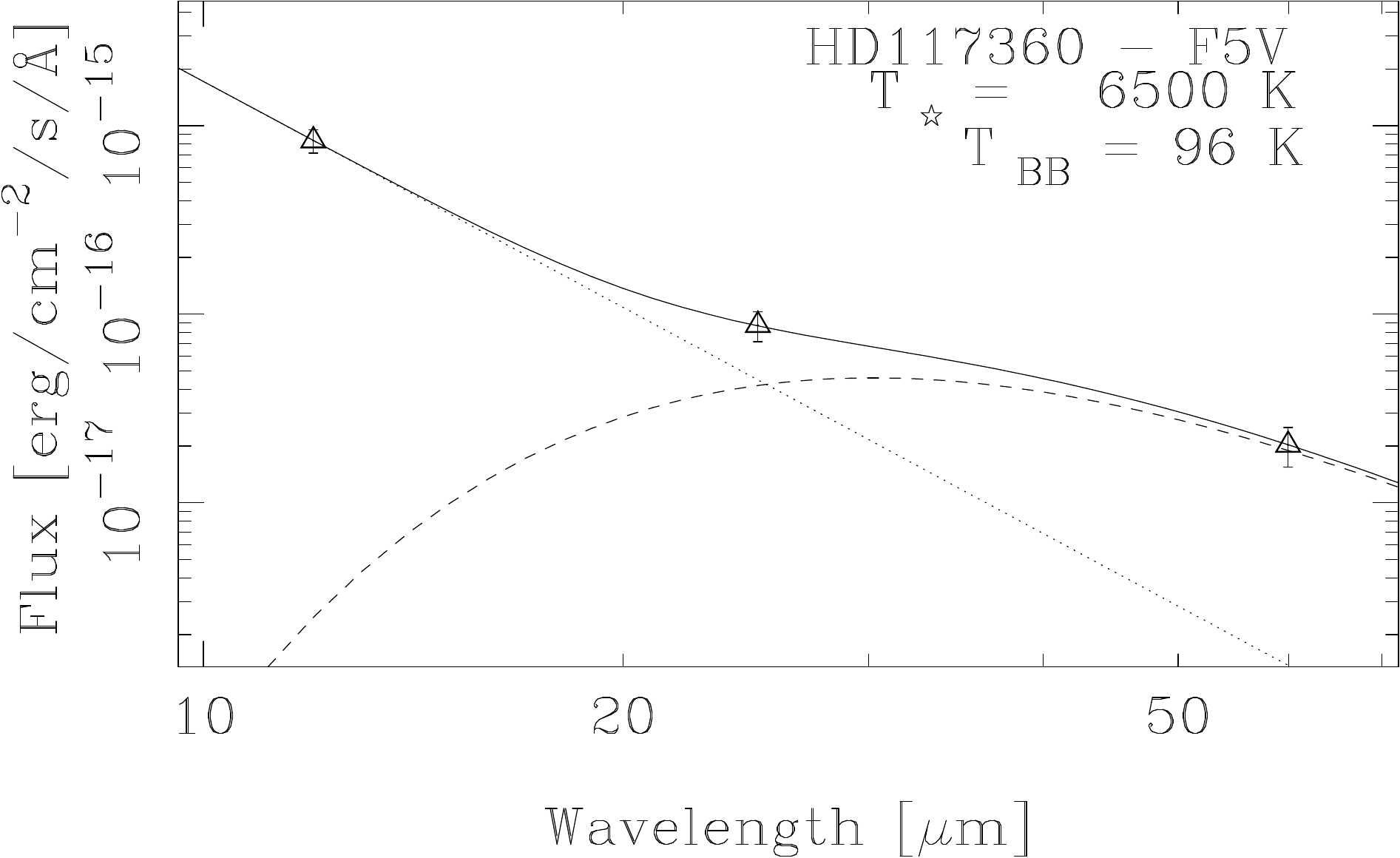}
\includegraphics[scale=0.28,angle=0,clip=true] {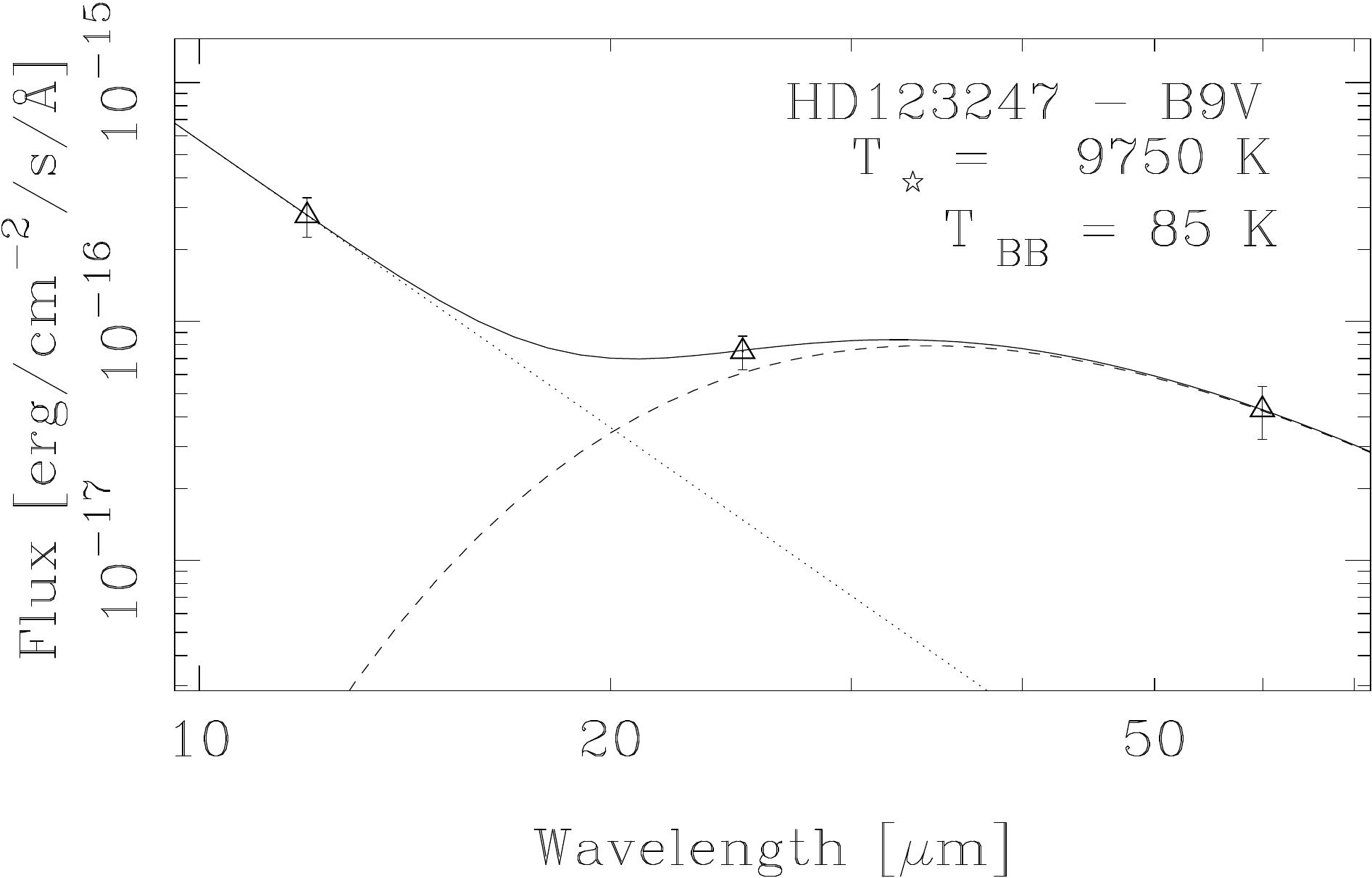}
\includegraphics[scale=0.28,angle=0,clip=true] {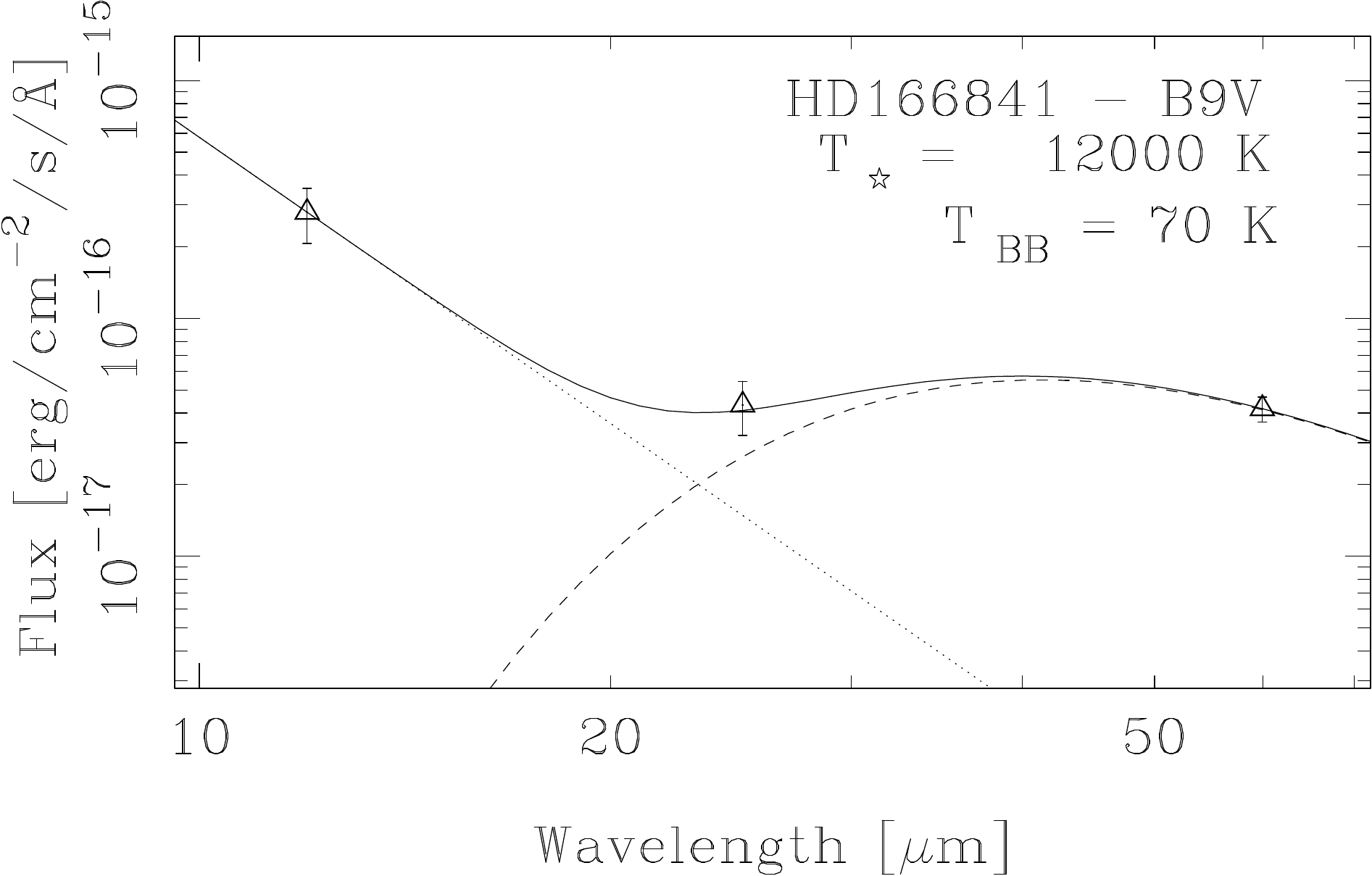}
\includegraphics[scale=0.28,angle=0,clip=true] {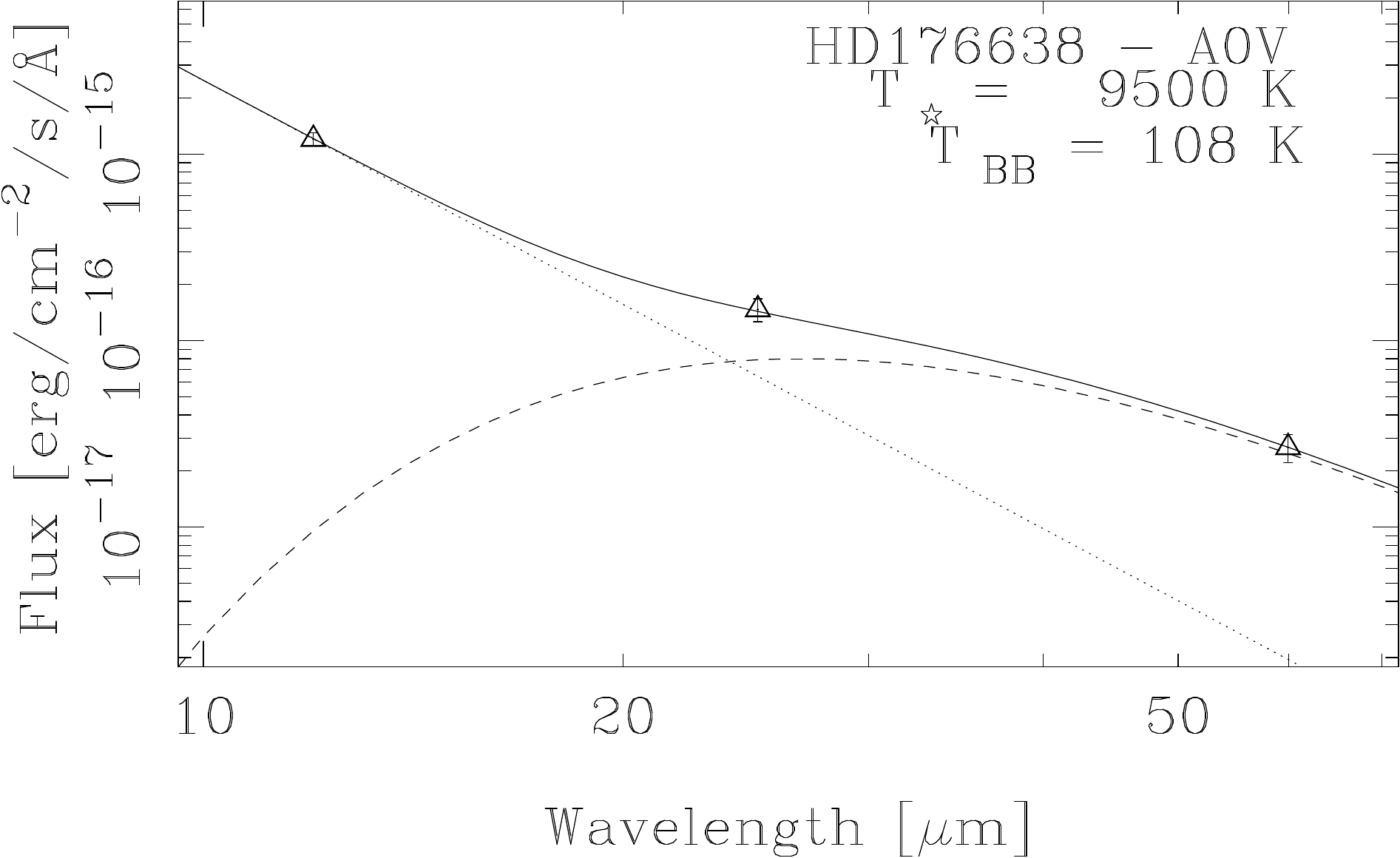}
\includegraphics[scale=0.28,angle=0,clip=true] {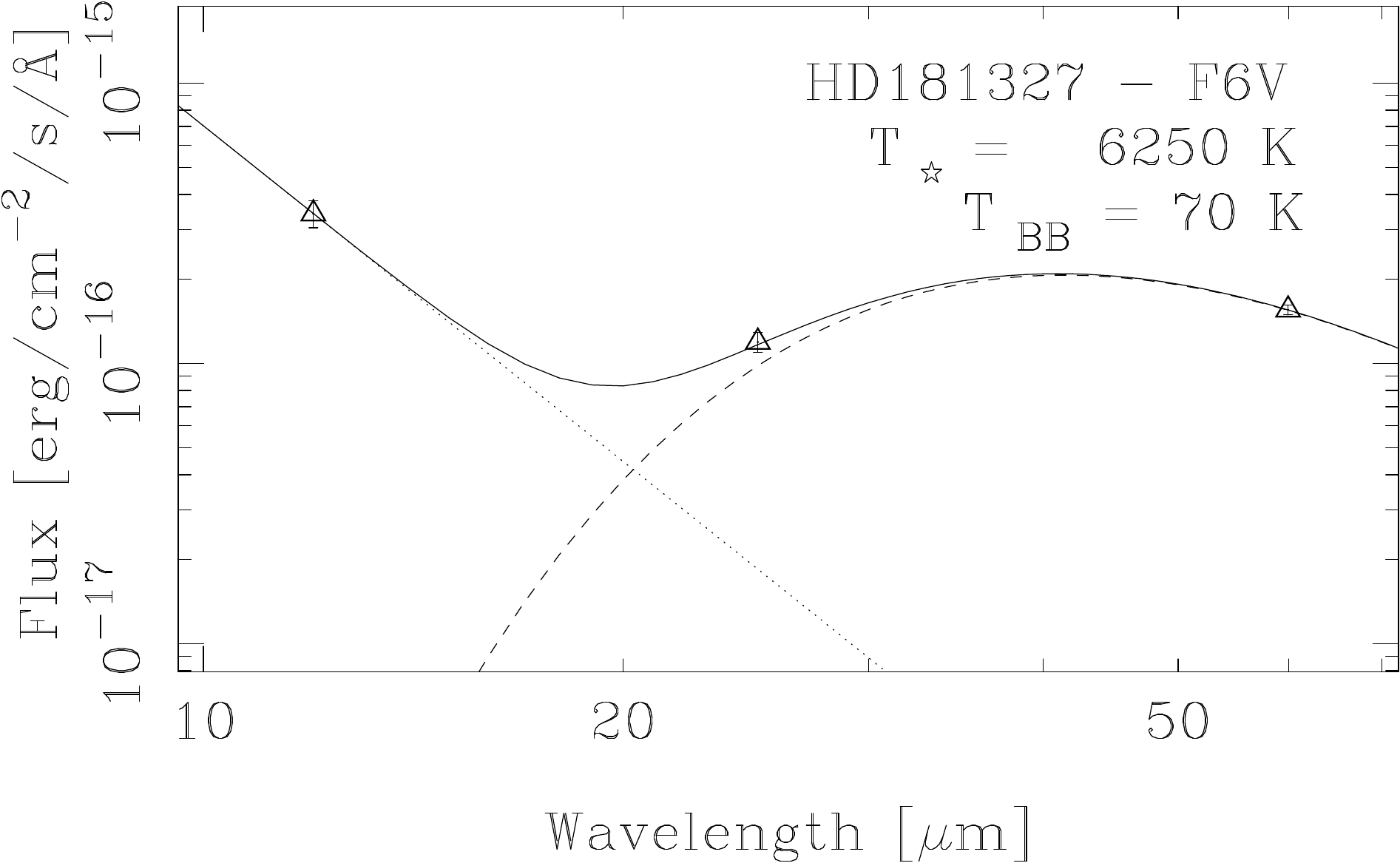}
\includegraphics[scale=0.28,angle=0,clip=true] {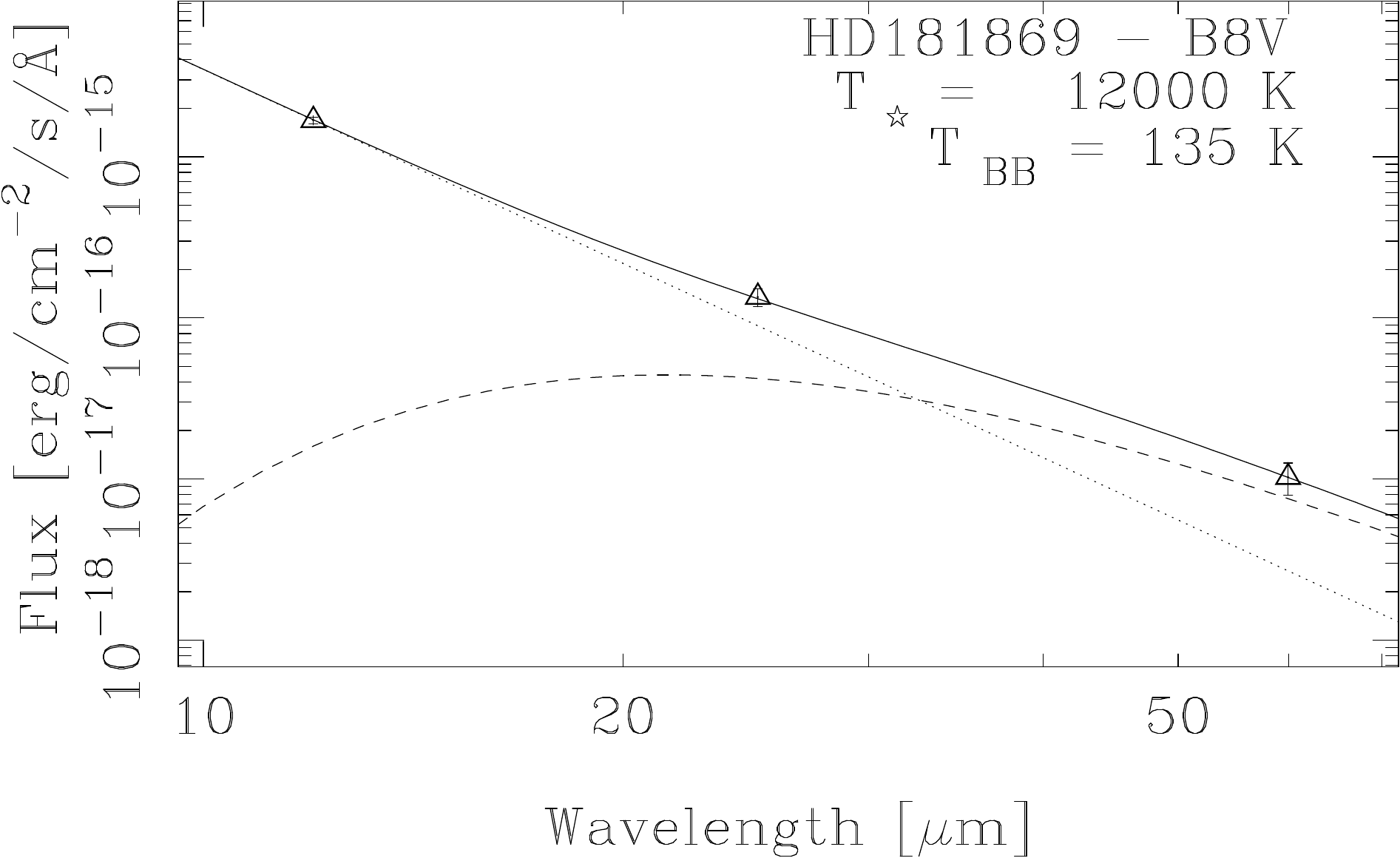}
\includegraphics[scale=0.28,angle=0,clip=true] {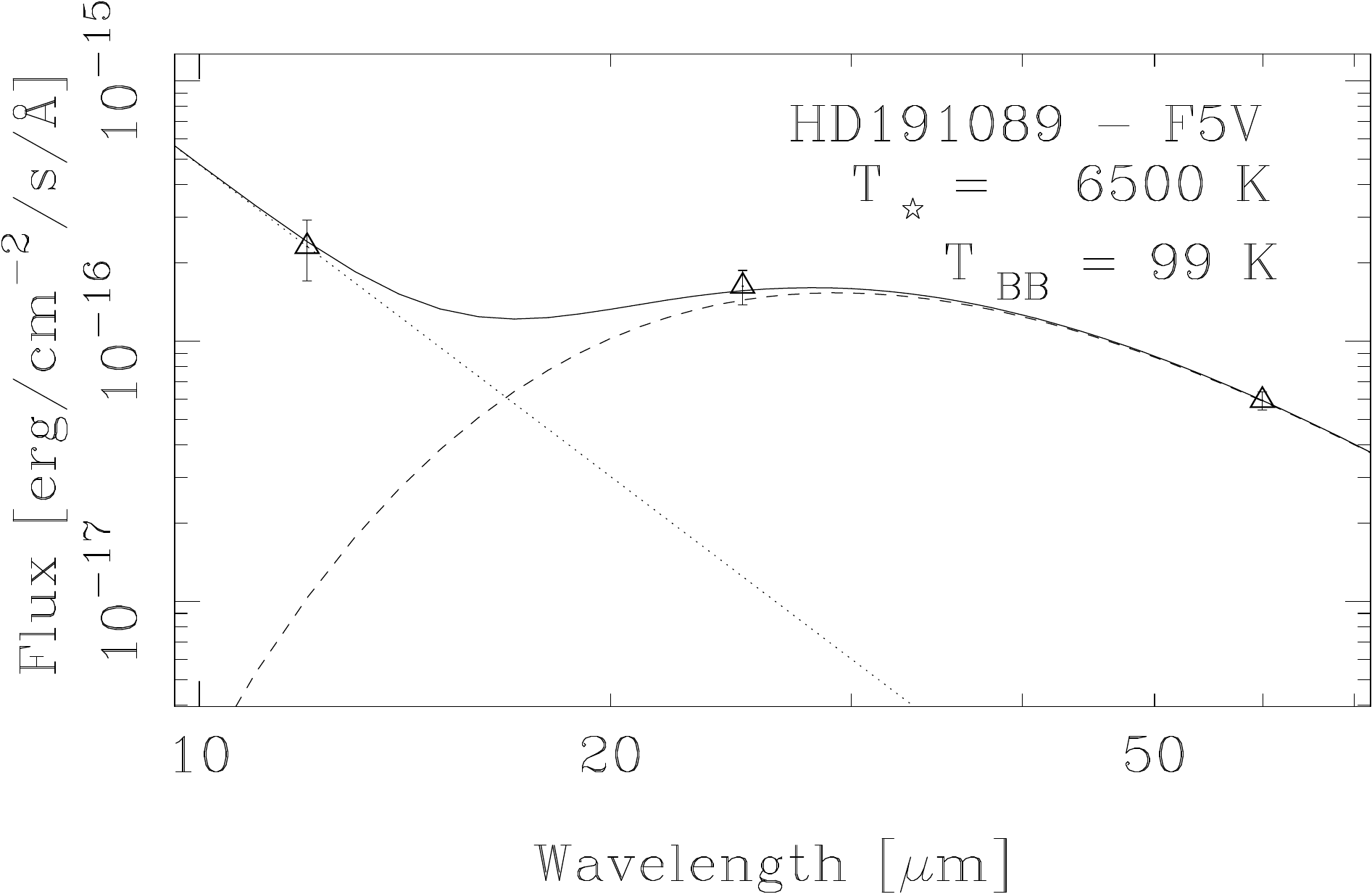}
\includegraphics[scale=0.28,angle=0,clip=true] {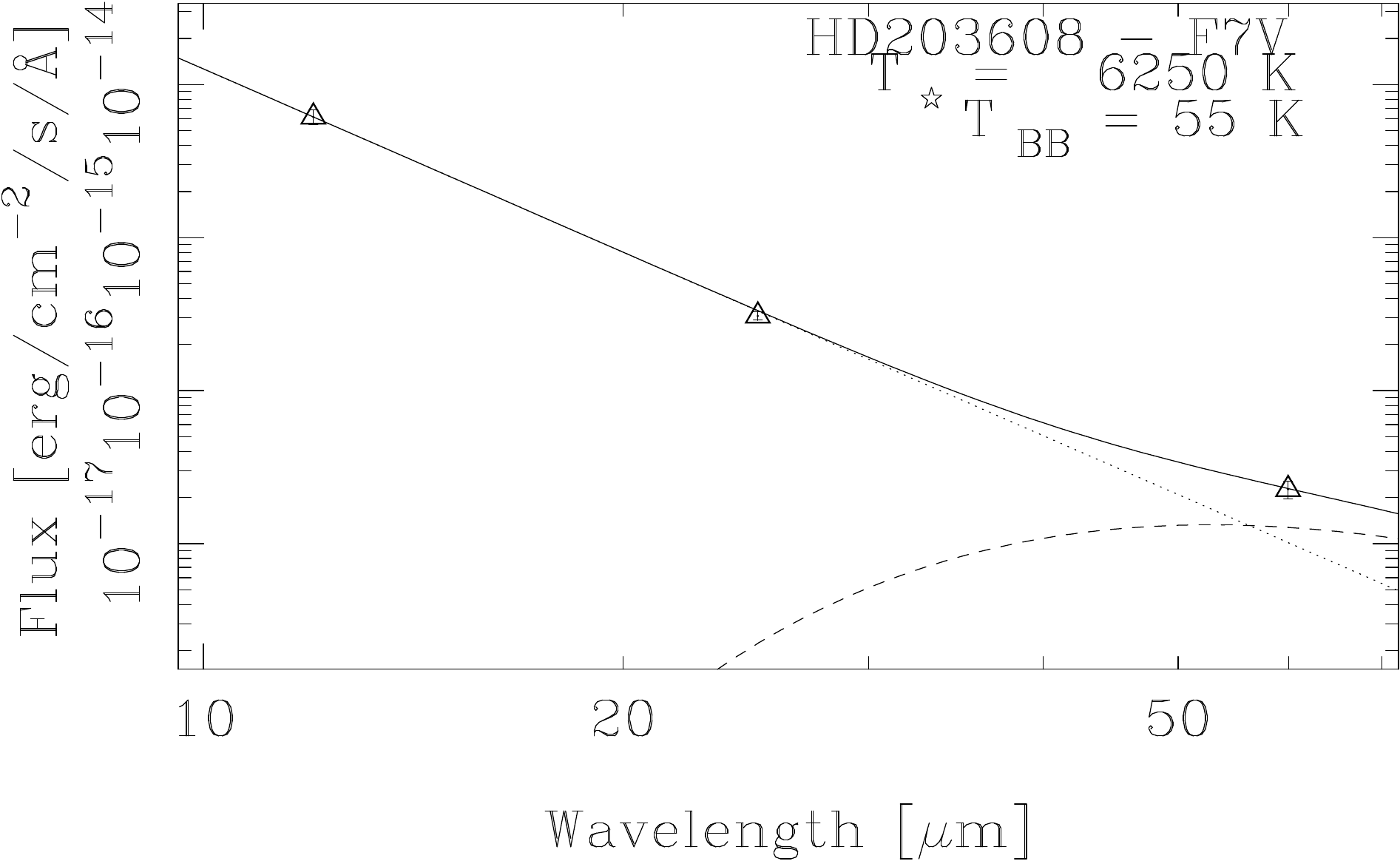}
 
\caption{Dust black-body fits to the infrared excesses used in the
  derivation of L$_{\rm IR}$/L$_{\rm \star}$ values for stars with
  different spectral types in our sample. The observed {\em IRAS}
  fluxes, as listed in \citet{mann98}, are plotted in blue with error
  bars denoting the 1$\sigma$ photometric errors. The dotted line in
  red corresponds to the reference stellar SED from
  \citet{castelli2004}, normalised to the 12~$\mu$m flux. The dashed
  line represents the fitted black-body and the solid line represents
  the resulting SED (SED$_{\rm \small \star}$+SED$_{\rm \small Dust
    Black body }$) that provides a satisfactory fit to the data.  The
  stellar temperatures characteristic of the corresponding stellar
  model from the \citet{castelli2004} library and the temperature of
  the fitted blackbody are indicated in each plot.}\label{seds}
\end{figure*}

Figure~\ref{seds} shows the results from the SED fitting routine,
where both the observed SED and the fitted {\it{stellar + dust
    blackbody model}} model are plotted. L$_{\rm IR}$/L$_{\rm \star}$
values in our sample are found to range between 1.8$\times10^{-3}$ and
7.5$\times10^{-6}$. We estimate our method to be accurate to factors
of 2-3 by comparing our $L_{\rm IR}$/$L_{*}$ results with values
already published in the literature
\citep{sylvester00,moor2006,hales2014}.

Based on their dust fractional luminosities, these systems are
expected to be in a more advanced evolutionary stage than the sample
studied by \cite{dun97b}. All stars have {\em Hipparcos} distances
available \citep{Leeuwen2007}. Spectral standards of three spectral
types were also observed in order to allow for spectral classification
and removal of telluric absorption lines. Table~\ref{sample}
summarizes the properties of our sample, where we have listed the
properties of our observed target sample along with those of the 3
observed spectral standards (HR~5558, HR~5670 and HR~6045).

\section{Observations and data reduction}

\subsection{AAT/UCLES Observations}

Optical Echelle spectroscopy of the 16 debris disc host stars was
obtained using the UCL Echelle Spectrogaph (UCLES) at the 3.9 meter
Anglo-Australian Telescope (AAT). All observations were taken on the
night of May 20$^{th}$ 2000. A log of the observations is given in
Table~\ref{log}, including the slit-width and seeing values. The
31.6~g/mm grating was used in conjunction with the
MIT/LL~CCD. Observations were carried out with two wavelength settings
in order to cover the blue region of the spectra between
3834-5440~$\rm{\AA}$, and between 5550-9900~$\rm{\AA}$ in the red. The
red observations were optimized for the study of H$_{\alpha}$, He~{\sc
  i} and the Na~{\sc i} D-lines, whilst the blue exposures covered the
Ca~{\sc ii} K and H lines as well as several metallic lines in the
4000-5000~$\rm{\AA}$ region.  The spectral resolution, estimated from
measuring the $FWHM$ of Thorium-Argon arc lines, was 0.115~$\rm{\AA}$
($\sim 8.8~$km s$^{-1}$) at Ca~{\sc ii} K (3933.663 $\rm{\AA}$),
corresponding to a resolving power of $R=34200$.  At the location of
the Na~{\sc i} D-lines, in the red part of the spectrum
(5889.951~$\rm{\AA}$), the spectral resolution and resolving power
were found to be 0.171~$\rm{\AA}$ and $R=34600$ respectively.

\begin{table*}
\setlength{\tabcolsep}{1mm}
\renewcommand{\arraystretch}{0.5}
\scriptsize
 \begin{center} 
\begin{tabular}{@{}ccccccccccc}
\\
\hline
\\
Star          &Other  & Previous     &     V & (B-V)             &Distance            &       L$_{\rm IR}$/L$_{\star}$   & L$_{\star}$   & Age [Myr]      &            References  \\
              &Name  & Spectral Type &       &                   &  $(pc)$              &                                & $(L_{\odot})$   &                &            for Age            \\
\hline	        							       
\hline\\        						       
	        							       
HD~53842      &  &    F5V       & 7.46  & 0.460          & $57.3 $& 2.6$\times10^{-4}$ &  5.0      & -        &                      \\
HD~61950      &  &    B8V       &  6.89 &-0.028          &$362.3 $& 1.4$\times10^{-4}$ &  1121.8   & -        &                           \\
HD~75416      &$\eta$~Cha  &    B8V       & 5.46  &-0.094& $96.9 $& 1.8$\times10^{-5}$ &  206.8    & 8        &0                   \\
HD~80950      &  &    A0V       &  5.87 &-0.016          & $80.8 $& 6.6$\times10^{-5}$ &  48.8     & 80       &1            \\
HD~88955      &q Vel  &    A1V/A2V    & 3.84  &0.063    &  $31.5 $& 7.5$\times10^{-6}$ &  41.7     &300       &1          \\
HD~91375      &  &    A1V       & 4.72  &0.042           & $79.4 $& 1.3$\times10^{-5}$ &  123.4    &265       &1       \\
HD~99211      &$\gamma$~Crt  &    A9V  & 4.08  &0.216   &  $25.7 $& 1.1$\times10^{-5}$ &  18.8     &600       &2          \\ 
HD~110058     &  &    A0V       & 7.98  &0.148           &  $107 $& 2.0$\times10^{-3}$ &  33.0     & 17   &5             \\
HD~117360     &S~Cha  &   F5V/F6V     & 6.48  &0.480     &  $35.1$& 1.5$\times10^{-4}$ &  5.4      & -        &               \\
HD~123247     &  & B9.5V/B9V     & 6.43  &0.000          & $101.1$& 2.4$\times10^{-4}$ &  52.3     & -        &                 \\
HD~166841     &  &   B8V/B9V   & 6.32  &-0.030           & $213.7$& 1.1$\times10^{-4}$ &  450.8    &-         &                \\
HD~176638     &$\zeta$~CrA  &   B9V/A0V  & 4.75  &-0.027 &  $56.3$& 5.1$\times10^{-5}$ &  64.4     & -        &               \\
HD~181327     &  &   F5V/F6V     &7.04   & 0.480         &  $50.6$& 2.4$\times10^{-3}$ &  4.1      &$12_{-4}^{+8}$&3         \\
HD~181869     &$\alpha$~Sgr   &    B8V      & 3.95&-0.083&  $52.1$& 7.9$\times10^{-6}$ &  161.4    &110$^{3}$     &1               \\
HD~191089     &  &    F5V       & 7.17  & 0.480          &  $53.5$& 1.8$\times10^{-3}$ &  3.5      &$12_{-4}^{+8}$&3             \\
HD~203608     &$\gamma$~Pav  &   F7V/F6V& 4.22  & 0.494&     $9.2$& 1.0$\times10^{-5}$ &  2.4      &10$^{3}$   &4             \\
\\	         	       	                  					                    
HR~5558       &$\beta$~Cir  &    A0V       & 5.32  & 0.044& $75.8$&  -   &       -   &  -    &  -           &             \\
HR~5670       &             &    A3V       & 4.06  & 0.100& $29.6$& -    &       -   &  -    &  -           &             \\
HR~6045       &$\theta$~Nor &    B8V       & 5.12  &-0.104& $90.4$& -    &       -   &  -    &  -           &              \\

\hline
\end{tabular}
\\
\caption{
\small Debris disc host stars observed along with spectral standards
used for spectral classification. Spectral types are from Michigan
Spectral Catalogue \citep{houk88} and distances are derived from
parallax measures listed in the {\em Hipparcos} catalogue
\citep{Leeuwen2007}. B-V colours are taken from the Simbad
database. The L$_{\rm IR}$/L$_{\star}$ values were derived using the
25~$\mu$m and 60~$\mu$m {\em IRAS} excess fluxes.  L$_{\star}$ is the
luminosity of the \citet{castelli2004} stellar model, normalised to
the 12~$\mu$m flux. L$_{\rm IR}$ is the bolometric luminosity of the
black-body that fits the infrared excesses.
}

References: (0) \cite{mamajek1999}, (1) \cite{rieke04}, (2) \cite{song01}, (3) \cite{moor2006}, (4) \cite{bryden06}, (5) \cite{kasper2015} .
\label{sample}
\end{center} 
\end{table*} 



%
%
%
%
%




%
%

\begin{table*}
\setlength{\tabcolsep}{1mm}
\renewcommand{\arraystretch}{0.5}
\scriptsize
 \begin{center} 
\begin{tabular}{@{}ccccccccc}
\\
\hline
\\
Object      & RA (J2000) &Dec (J2000)&  Airmass&Seeing&Exposure& Central&      Slitwidth&Slitlength\\
   &    &  &  &(arcseconds)&(s)&  Wavelength ($\rm{\AA}$) &    (arcseconds)&(arcseconds)\\
\hline	      							       
\hline\\     
HD~53842     & 06:46:00.45&-83:59:35.2&   1.780 &  1.5 &1000& 4400.03   &        1.47&  6.96 \\
   & &&   1.803 &  1.5 &1000& 4400.03   &        1.47&  6.96 \\
   & &&   1.832 &  1.5 &1000& 4400.03   &        1.47&  6.96 \\
   & &&   2.249 &  2.5 & 400& 6699.95   &        1.47&  5.6 \\
\\
HD~61950     & 07:36:16.97&-69:03:29.6&   1.384 &  1.5 & 750& 4400.03   &        1.47&  6.96 \\
&&&   1.413 &  1.5 & 750& 4400.03   &        1.47&  6.96 \\
&&&   3.699 &  2.5 & 350& 6699.95   &        1.47&  5.6 \\
\\
HD~75416     & 08:41:14.77&-78:57:55.3&   1.496 &  1.5 & 250& 4400.03   &        1.47&  6.96 \\
&&&   1.501 &  1.5 & 250& 4400.03   &        1.47&  6.96 \\
&&&   2.348 &  2.5 & 200& 6699.95   &        1.47&  5.6 \\
\\
HD~80950     & 09:17:24.97&-74:44:09.9&   1.384 &  1.5 & 500& 4400.03   &        1.47&  6.96 \\
&&&   2.447 &  2.5 & 200& 6699.95   &        1.47&  5.6 \\
\\
HD~88955     & 10:14:44.29&-42:07:21.1&   1.080 &  1.5 & 100& 4400.03   &        1.47&  6.96 \\
&&&    2.438 &  2.5 &  50& 6699.95   &        1.47&  5.6 \\
\\
HD~91375     & 10:30:18.21&-71:59:40.1&   1.353 &  1.5 & 200& 4400.03   &        1.47&  6.96 \\
&&&     2.195 &  2.5 &  60& 6699.95   &        1.47&  5.6 \\
\\
HD~99211     & 11:24:53.33&-17:41:05.1&   1.043 &  1.5 & 150& 4400.03   &        1.47&  6.96 \\
&&&     2.188 &  2.5 &  40& 6699.95   &        1.47&  5.6 \\
\\
HD~110058    & 12:39:46.24&-49:11:54.5&   1.475 &  1.5 &1000& 4400.03   &        1.47&  6.96 \\
&&&   1.471 &  1.5 &1000& 4400.03   &        1.47&  6.96 \\
&&&   1.473 &  1.5 &1000& 4400.03   &        1.47&  6.96 \\
&&&   1.345 &  2.5 &1000& 6699.95   &        1.47&  5.6 \\
\\
HD~117360    & 13:33:10.61&-77:34:13.7&   1.446 &  2.5 & 800& 4400.03   &        1.47&  6.96 \\
&&&   1.446 &  2.5 & 800& 4400.03   &        1.47&  6.96 \\
&&&   1.625 &  3.0& 400& 6699.95   &        1.47&  5.6 \\
\\
HD~123247    & 14:07:40.85&-48:42:14.8&   1.048 &  2.5 &1000& 4400.03   &        1.47&  6.96 \\
\\
HD~166841    & 18:17:59.57&-68:13:46.0&   1.318 &  2.3 & 800& 4400.02   &        1.47&  7.0 \\
&&&   1.256 &  3.0 & 400& 6699.95   &        1.47&  5.6  \\
\\
HD~176638    &19:03:06.95&-42:05:39.6&    1.075 &  2.3 & 250& 4400.02   &        1.47&  7.0 \\
&&&   1.048 &  3.0 &  80& 6699.95   &        1.47&  5.6 \\
\\
HD~181327    & 19:22:58.63&-54:32:17.2&   1.127 &  2.0 & 900& 4400.02   &        1.47&  7.0 \\
&&&    1.150 &  2.0 & 900& 4400.02   &        1.47&  7.0 \\
&&& 1.119 &  2.3 & 400& 6699.95   &        1.47&  5.6 \\
\\
HD~181869    & 19:23:53.25&-40:36:56.3&   1.024 &  2.3 & 100& 6699.95   &        1.47&  5.6 \\
&&&    1.114 &  2.0 & 100& 4400.02   &        1.47&  7.0 \\
&&&    1.039 &  2.3 &  40& 6699.95   &        1.47&  5.6 \\
\\
HD~191089    & 20:09:05.54&-26:13:23.1&   1.013 &  2.3 &1000& 4400.02   &        1.47&  7.0 \\
&&&   1.005 &  2.3 &1000& 4400.02   &        1.47&  7.0 \\
&&&   1.470 &  2.3 & 400& 6699.95   &        1.47&  5.6 \\
&&&   1.087 &  2.3 & 400& 6699.95   &        1.47&  5.6 \\
\\
HD~203608    & 21:26:25.91&-65:21:56.9&  1.207  &  2.0 &  200.0&  4400.02    &      1.47&   7.0 \\
 &&&      1.357 &  2.3 & 100& 6699.95   &        1.47&  5.6 \\
\\
HR~5558      & 14:55:44.86&-33:51:19.9&   1.006 &  2.5 & 500& 4400.03   &        1.47&  6.96 \\
&&&    1.164 &  3.0 & 120& 6699.95   &        1.47&  5.6 \\
\\
HR~5670      & 15:17:30.36&-58:48:04.9&   1.134 &  2.5 & 200& 4400.03   &        1.47&  6.96 \\
&&&   1.225 &  3.0 &  40& 6699.95   &        1.47&  5.6 \\
\\
HR~6045      & 16:15:15.23&-47:22:17.4&   1.085 &  2.5 & 400& 4400.03   &        1.47&   6.96\\
&&&    1.048 &  2.5 & 100& 6699.95   &        1.47&   5.6\\

\hline
\end{tabular}

\caption{Log of UCLES observations.}\label{log}

\end{center}
\end{table*}
\normalsize

 Overscan correction, dark subtraction and flat-fielding were performed
using the Image Reduction and Analysis Facility (IRAF \footnote{\tt
http://www.iraf.noao.edu}) package developed by the National Optical
Astronomy Observatory (NOAO). When more than one exposure for the same
star was available, the different frames were coadded and the average
taken. At this stage the `crreject' task was used in order to remove
cosmic rays.
Echelle spectra were extracted following standard echelle reduction
procedures using the different tasks of the Echelle package in
IRAF. These include order tracing, extracting and blaze correcting
each order. Wavelength calibration was performed using reference arc
lines from a Thorium-Argon lamp. Each order was wavelength calibrated
individually, producing a dispersion accuracy always better than
0.004~$\rm{\AA}$.


\subsection{Magellan/MIKE Observations}

Follow up observations of the target HD~110058 were obtained using the
Magellan Inamori Kyocera Echelle (MIKE) spectrograph
\citep{bernstein2003}, mounted on the Magellan II telescope. In order
to characterize the distribution of interstellar material in the
surroundings of HD~110058, three stars located in the same direction
and within a small range of distances from the main target were
observed.

All observations were carried out in the night of 2008 July 4 under
clear sky conditions and with seeing between 0.9 and
1.2~arcseconds. The summary of the observations together with the
target properties are shown in Table~\ref{logmike}. MIKE allows to
obtain both the red and blue side of the spectrum simultaneously,
covering form 3350 to 9150 $\AA$. The 0.7x5 ~arcsec$^2$ slit was used
which provided spectral resolutions of 42000 and 32000 in the blue and
red parts of the spectra, respectively (as measured by fitting the
FWHM of the Thorium-Argon lamp lines). All targets are bright (V$<$8), so
signal-to-noise ratios (SN) higher than 15 can be obtained in both the
blue and red arms in only a few minutes of total integration time
(listed in the last two columns of Table~\ref{logmike}).

Data reduction was performed using the CarPy MIKE data reduction
package \citep{kelson2003}, which performs the standard steps for
echelle reduction i.e. overscan subtraction, order tracing, sky
subtraction, extraction, wavelength calibration. The pipeline also
combines multiple frames of the same target and to produce final
combined spectra.

\begin{table*}
\setlength{\tabcolsep}{1mm}
\renewcommand{\arraystretch}{0.5}

 \begin{center} 
\begin{tabular}{@{}ccccccccc}
\\
\hline
\\
Object      & RA (J2000) &Dec (J2000)&  Spectral &V & Distance& Angular& Integration& SN \\
            &             &          &   Type    &  &(pc)     &Separation& Time (s)&  Blue/Red \\
\hline	      							       
\hline   
HIP~61557&12:36:46.60&-50:20:07.42&A0V&6.39&102&1.2$^{\circ}$ & 198 & 40/30\\
HD~110058&12:39:46.24&-49:11:55.54&A0V&7.98&107&0$^{\circ}$   & 480 & 25/21\\
HIP~61342&12:34:08.84&-50:03:36.95&A3V&7.23&115&1.3$^{\circ}$ & 120 &17/15\\
HIP~60360&12:22:36.52&-49:19:34.07&A2IV/V&7.94&123&2.8$^{\circ}$ & 207&17/15\\
\hline
\end{tabular}

\caption{Summary of MIKE observations. Spectral types are from the
  Michigan Spectral Catalog. Distances are derived from {\em
    Hipparcos} parallaxes. Column 6 denotes the angular separation from
  the direction of HD~110058.}\label{logmike}
 
\end{center}
\end{table*}
\normalsize

\section{Spectral classification and stellar parameters}

In order to check the accuracy of previous spectral classifications of
our sources, several regions of the blue parts of the acquired spectra
were compared to those of the observed standards (as listed in
Table~\ref{sample}) and also to archive spectra from the UVES Paranal
Observatory Project \citep[UVES POP; ][]{bagnulo03}. The latter
provides a library of high-Resolution spectra of stars across the
Hertzsprung-Russell Diagram at a resolution of $R\sim 80000$, which
were degraded to the resolution of our UCLES spectra. Several regions
of the 4000-5000~$\rm{\AA}$ part of the observed spectra of the stars
were compared by eye to the spectra of the nearest spectral types. In
some cases the template spectra were artificially broadened in order
to match the rotational velocity of the star to be classified. No
discrepancies between the Michigan spectral types of the target stars
and their corresponding spectral templates were found, leading us to
conclude that all of our stars have their previous Michigan Spectral
Catalog classification confirmed. In cases when the Michigan Spectral
Catalog had an uncertainty of the order of 1 or 2 spectral
sub-classes, our method was able to discriminate and associate the
star with one of those spectral types, as listed in Table~\ref{ew}.

In the case of the five F-type stars (and the A9 star HD~99211), whose
spectra are rich in metallic species, radial velocities (V$_{\odot}$)
were derived by cross-correlating the observed spectra with template
spectra for stars of similar spectral type from the UVESPOP database,
for which radial velocities had already been measured. The FXCORR task
in IRAF was used for this purpose, which allows one to cross-correlate
several echelle orders simultaneously, providing a very accurate
determination of V$_{\odot}$. This is reflected in the small error
estimates presented in Table~\ref{ew}, which correspond to the
standard deviation of the velocities obtained in the different
orders. For the case of the late B-type and early A-type stars, these
stars have intrinsically fewer metallic lines that can be used for
radial velocity estimations. Therefore the most prominent photospheric
lines of Ca~{\sc ii}, Mg ~{\sc ii} and Fe~{\sc i} were used for
performing Gaussian fitting to the centres of the photospheric
lines. The heliocentric V$_{\odot}$ values listed in Table~\ref{ew}
for these stars correspond to the mean obtained using the different
lines. The Mg~{\sc ii} doublet at 4481.13~$\rm{\AA}$ and 4481.33
~$\rm{\AA}$ is unresolved in the case of the B- and early A-type
stars, and therefore was not used for radial velocity
determination. Earth radial velocity corrections were applied to each
spectral range, in order to obtain heliocentric velocities (obtained
using the Radial Velocity (RV) Starlink package).

Projected rotational velocities ($v\,\sin{i}$) were derived using the
STAROT package within DIPSO, which allows to artificially broaden the
spectrum of a stellar template to match the spectra of the observed
star. Template spectra with previously measured $v\,\sin{i}$ values
for each spectral type were taken from the UVESPOP database, by
selecting the stars with the lowest rotational velocities. Artificial
broadening was then applied until a satisfactory fit to the observed
spectra was achieved. Derived rotational velocities for our target
stars are presented in column 3 of Table~\ref{ew}.

\begin{table*}
\setlength{\tabcolsep}{1mm} \renewcommand{\arraystretch}{0.5}
\scriptsize
 \begin{center} 
\begin{tabular}{@{}ccccc}
\\
\hline
\\
             &               &$v\,\sin{i}$ & V$_{\odot}$  \\
Star          &Spectral Type &             &              \\
             &               & (km s$^{-1}$)& (km s$^{-1}$)          \\

\hline	      							       										  
\hline\\      						       											  
	      							       										  
HD~53842      &    F5V         & 24$\pm2$           &  10 $\pm1$     \\     
HD~61950      &    B8V         & 240 $\pm9$         & 20$\pm5$     \\     
HD~75416      &    B8V         & 290 $\pm2$         &15 $\pm5$     \\     
HD~80950      &    A0V         &80 $\pm4$            &14$\pm3$      \\     
HD~88955      &    A1V         &100 $\pm4$           &7$\pm3$       \\     
HD~91375      &    A1V         &15 $\pm2$            &8$\pm3$       \\     
HD~99211      &    A9V         &110 $\pm2$          &-1 $\pm2$     \\     
HD~110058     &    A0V         &180 $\pm4$          &12 $\pm3$     \\     
HD~117360     &    F5V         &11 $\pm2$           & -33 $\pm1$   \\     
HD~123247     &    B9V         &140 $\pm4$           & 7  $\pm3$    \\     
HD~166841     &    B9V         &245 $\pm8$          & 0  $\pm3$            \\     
HD~176638     &    A0V         &260 $\pm6$          &-27 $\pm3$          \\     
HD~181327     &    F6V         & 22  $\pm2$         & -1 $\pm1$     \\     
HD~181869     &    B8V         &  55$\pm9$          &-4  $\pm5$       \\     
HD~191089     &    F5V         & 33 $\pm2$          & -6 $\pm1$  \\     
HD~203608     &    F6V          & 3 $\pm1$          & -30 $\pm1$  \\     \\

HR~5558       &    A0V          &  180$\pm4$          & 1 $\pm3$  \\  
HR~5670       &    A3V          &  95 $\pm4$          & 9 $\pm2$  \\  
HR~6045       &    B8V          &  205$\pm5$          & 1 $\pm4$  \\

\hline
\end{tabular}

\medskip


\caption{Measured parameters of the debris disc host stars. Column 2 confirms the
previous spectral classification for all of our targets.  Columns 3
and 4 list the projected rotational velocities ($v\,\sin{i}$) and the
heliocentric radial velocities (V$_{\odot}$), respectively. The
uncertainties quoted for V$_{\odot}$ and $v\,\sin{i}$ correspond to
the dispersion obtained when deriving the quantities using the
different lines.}\label{ew}
\end{center}
\end{table*}
\normalsize

\section{Evidence for circumstellar gas}\label{csgas}

\begin{figure}
\begin{center}

\includegraphics[scale=0.3,angle=0,clip=true]{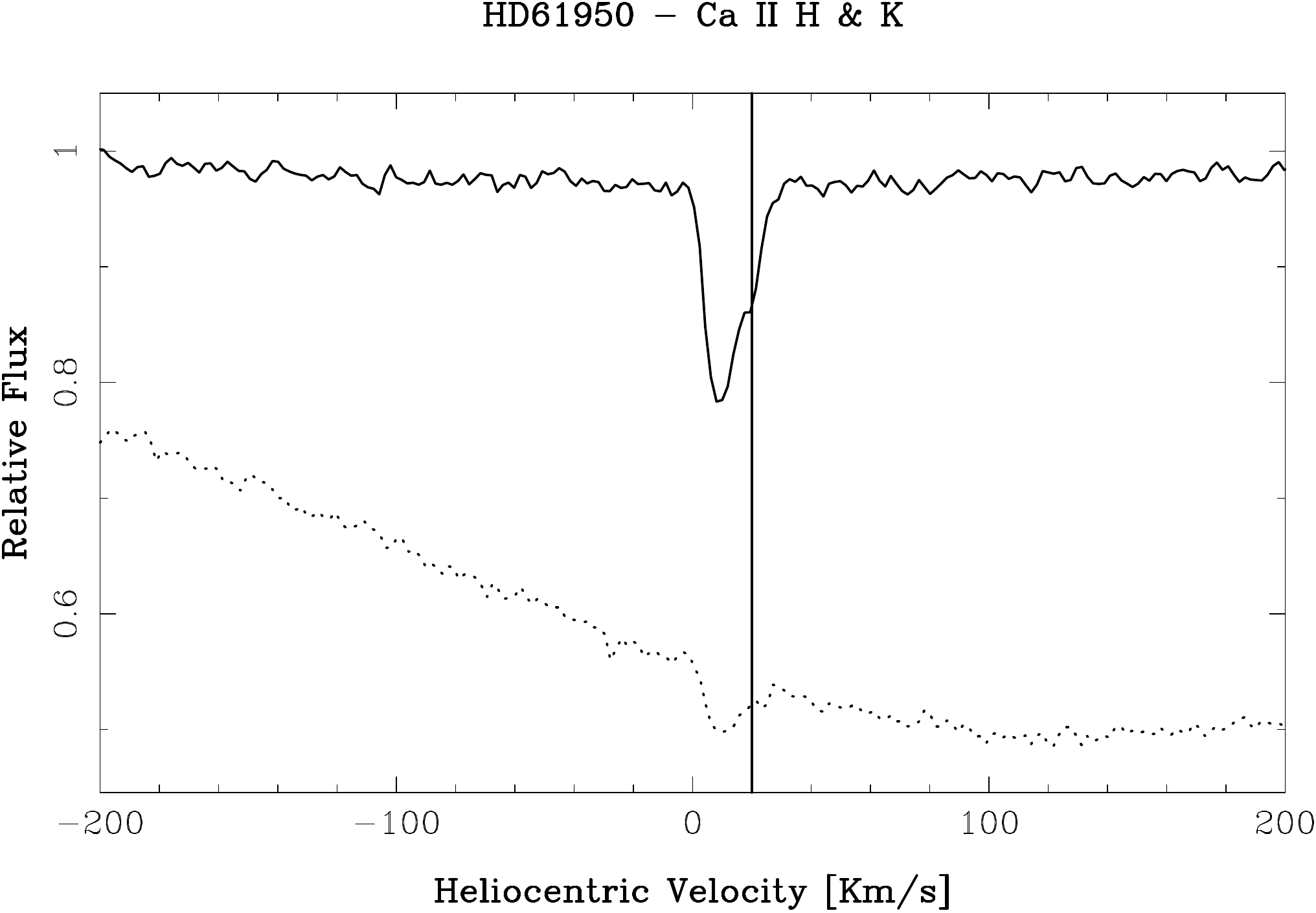}
\includegraphics[scale=0.3,angle=0,clip=true]{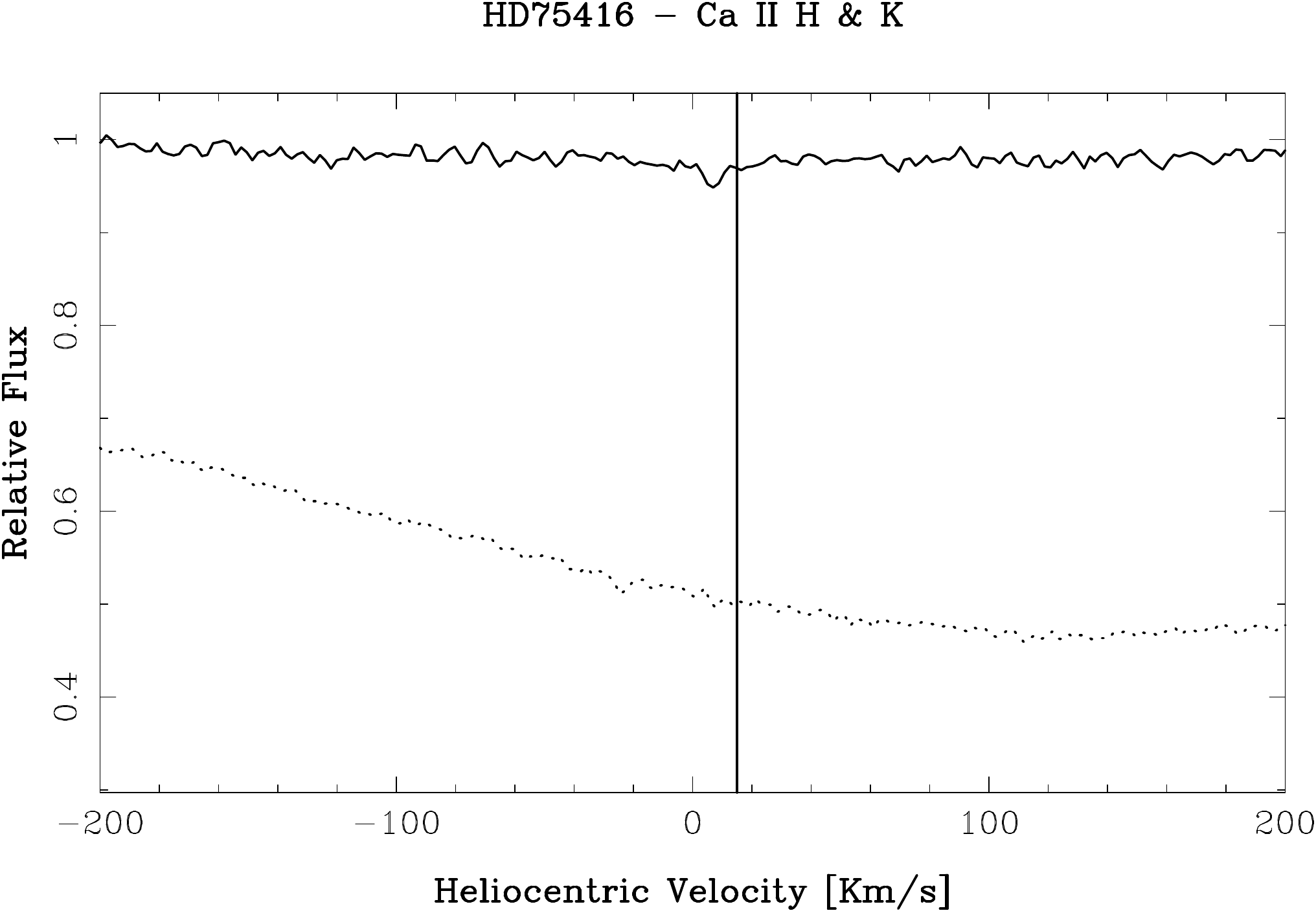}  
\caption{Heliocentric velocity plots of the narrow Ca~{\sc ii}~H
(dotted line) and K (solid line) absorption components present in the
spectra of HD~61950 and of HD~75416. The vertical lines denote the
radial velocity of the star. The H line spectra have been shifted
vertically for plotting purposes.}\label{fig3}
\end{center}
\end{figure}

\begin{figure}
\begin{center}
\includegraphics[scale=0.3,angle=0,clip=true]{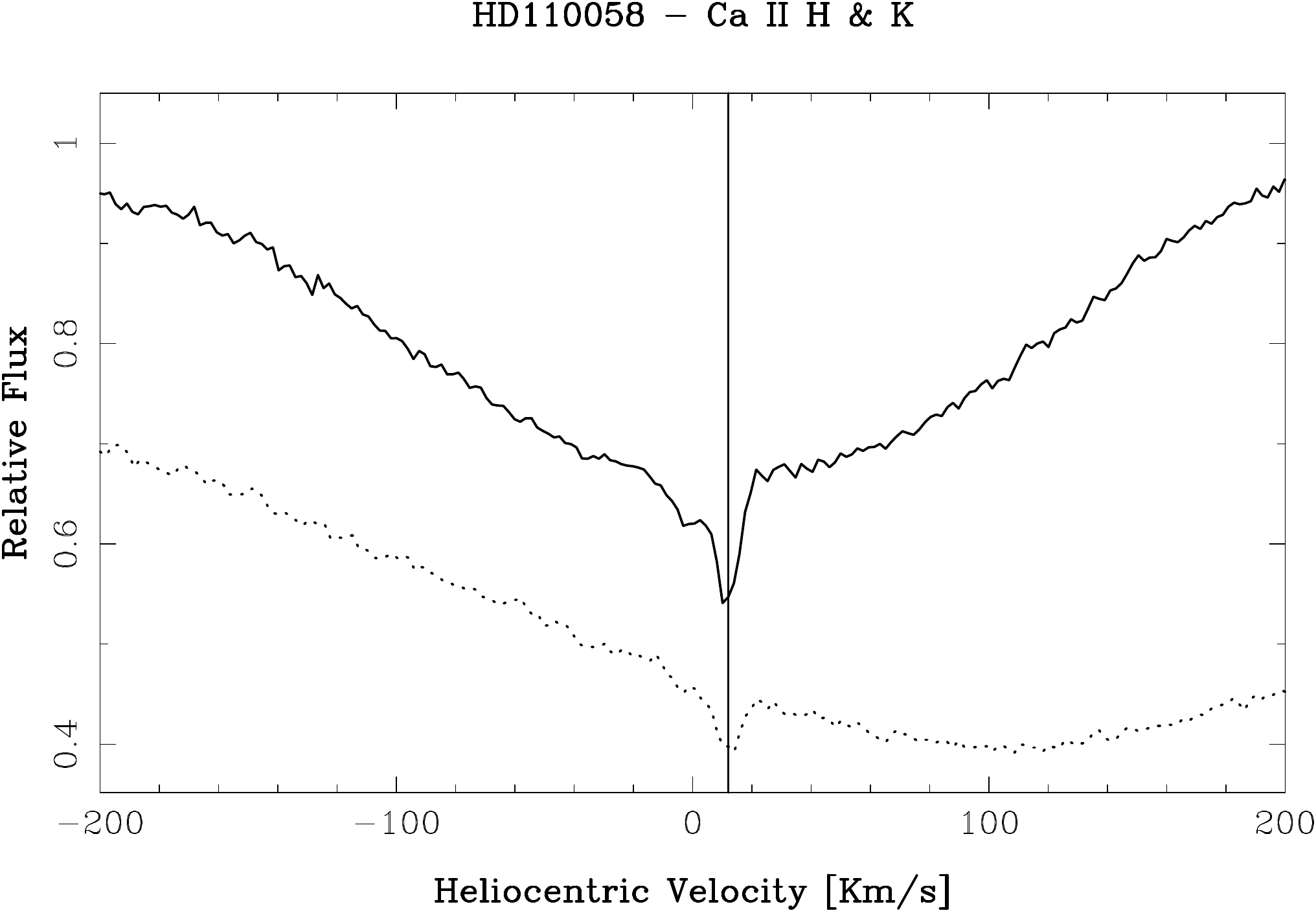}
\includegraphics[scale=0.3,angle=0,clip=true]{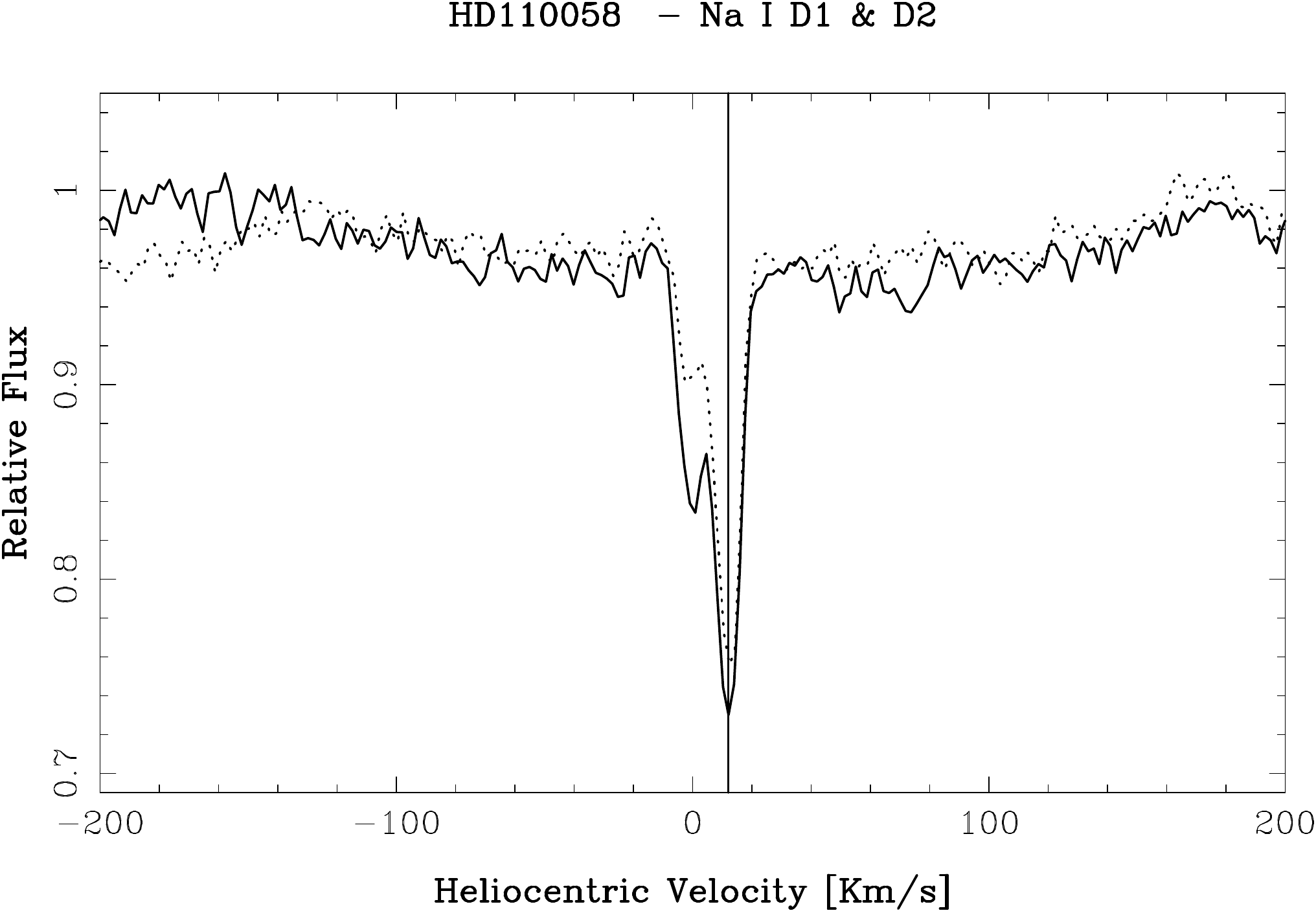}
%
\caption{Heliocentric velocity plots of narrow absorption features
seen towards HD~110058 in both the Ca~{\sc ii} H \& K lines (top panel) and in the
Na~{\sc i} D lines (bottom panel). The solid lines correspond to
the Ca~{\sc ii} K and Na~{\sc i} D$_2$ lines. The Na~{\sc i} region
of the spectra was divided by the spectrum of the B8 spectral
standard HR~6045 in order to remove telluric absorption features.}\label{fig4}
\end{center}
\end{figure}

Of the 16 targets observed, 4 are found to exhibit narrow absorption
features located near the centre of the photospheric Ca~{\sc ii}~H \&
K lines - these stars are HD~61950, HD~75416, HD~110058 and HD~166841.
Similar absorption features are seen in the Na~{\sc i} D lines, with
the exception of HD~61950 and HD~75416, for which only Ca~{\sc ii}
data were available. No narrow absorption components can be seen in
either the Ca~{\sc ii} K or Na~{\sc i} D line profiles of the
remaining stars of our sample. In addition, a narrow feature was also
detected in the Ca~{\sc ii}~H \& K spectrum of the spectral standard
HR~5558. Figure~\ref{fig3} shows the H \& K regions of HD~61950 and
HD~75416 in the heliocentric velocity frame, with the radial
velocities of the stars marked by a vertical line. Figure~\ref{fig4}
shows the Na~{\sc i} D and the Ca~{\sc ii}~H\& K lines of HD~110058,
HD~166841 and HR~5558 in the heliocentric velocity frame.  These
features are too narrow to be photospheric in nature and therefore
they must be produced by absorption of stellar light by either an
interstellar or circumstellar gas cloud.  There is a good
correspondence between the radial velocities of the components from
the different species, as is expected if both the calcium and sodium
are located in the same cloud. For both HD~110058 and HD~166841 there
is a strong absorption component that coincides, within the errors,
with the radial velocity of the star. HD~61950 and HD~75416 have
components that partially overlap the stellar radial velocity, while
in the case of HR~5558 the velocity of the absorption feature is
totally dissimilar to that of the star and is deemed to be
interstellar in origin (see \ref{hr5558}).\\





\begin{figure}
\begin{center}
 \vspace{1.5pt}                                                                                               
\includegraphics[scale=0.3,angle=0,clip=true]{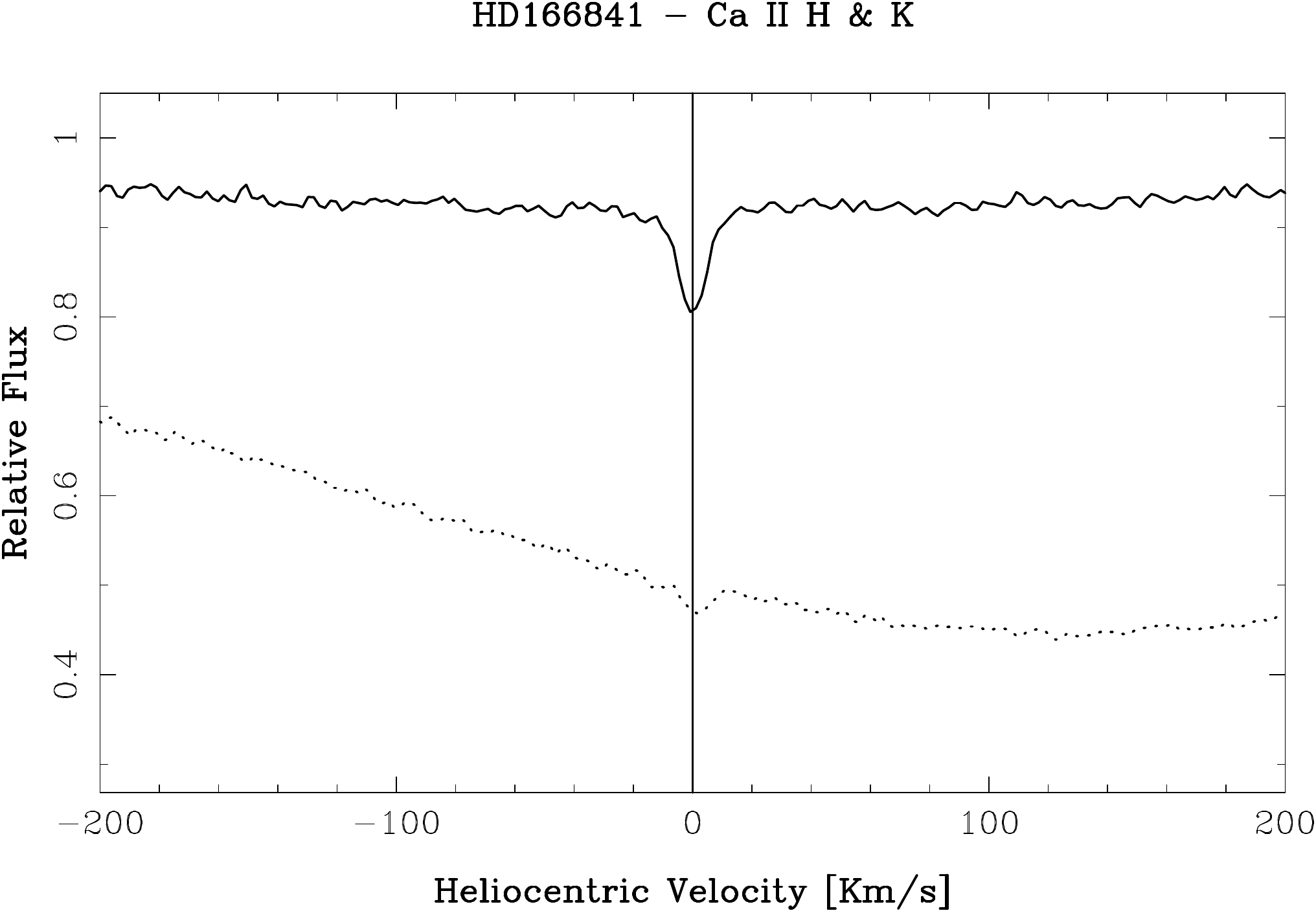}
\includegraphics[scale=0.3,angle=0,clip=true]{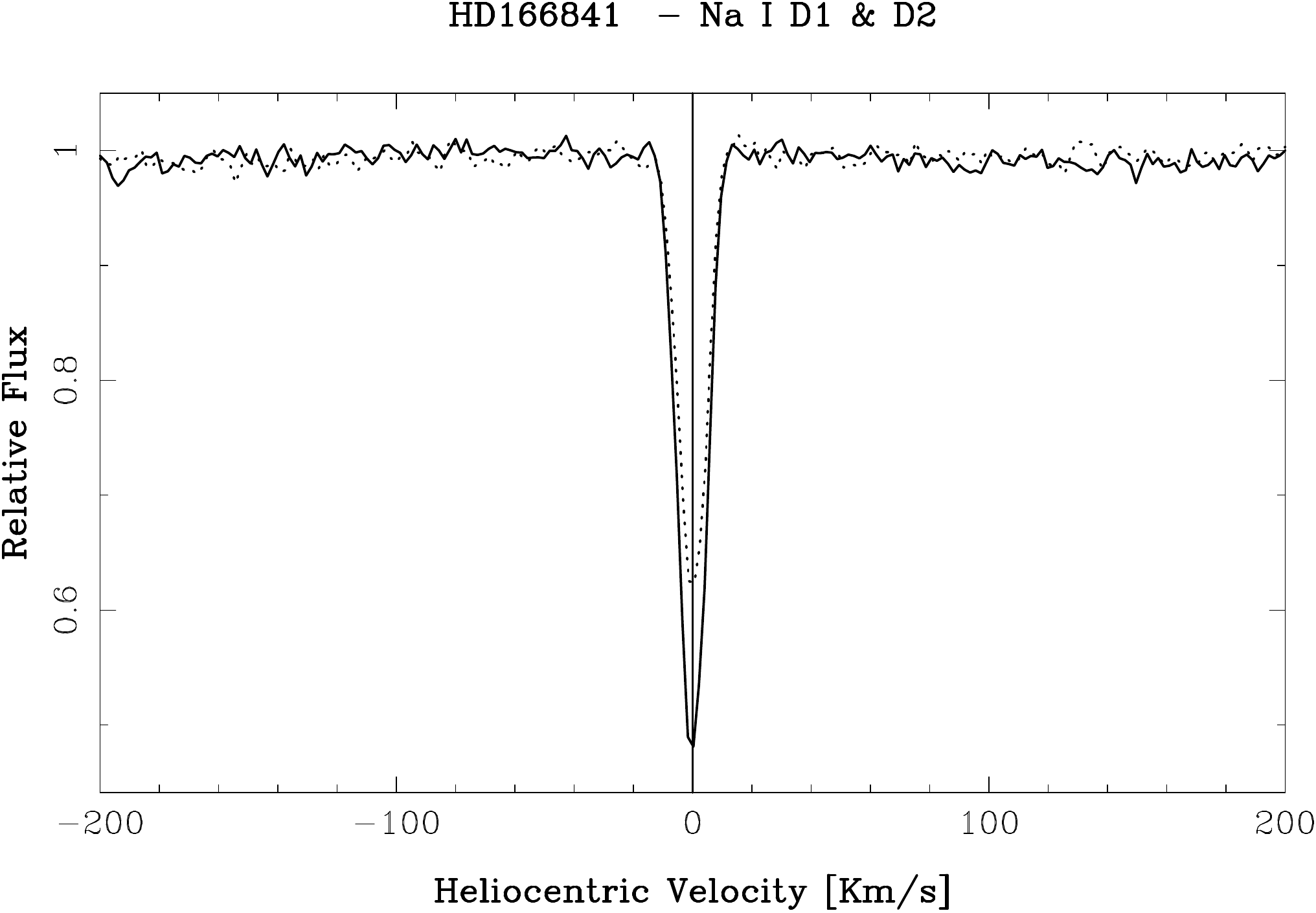}
 \vspace{1.5pt}                                                                                               

\caption{Same as Figure~\ref{fig4} for HD~166841}\label{fig4b}
\end{center}
\end{figure} 

\begin{figure}
\begin{center}
\includegraphics[scale=0.3,angle=0,clip=true]{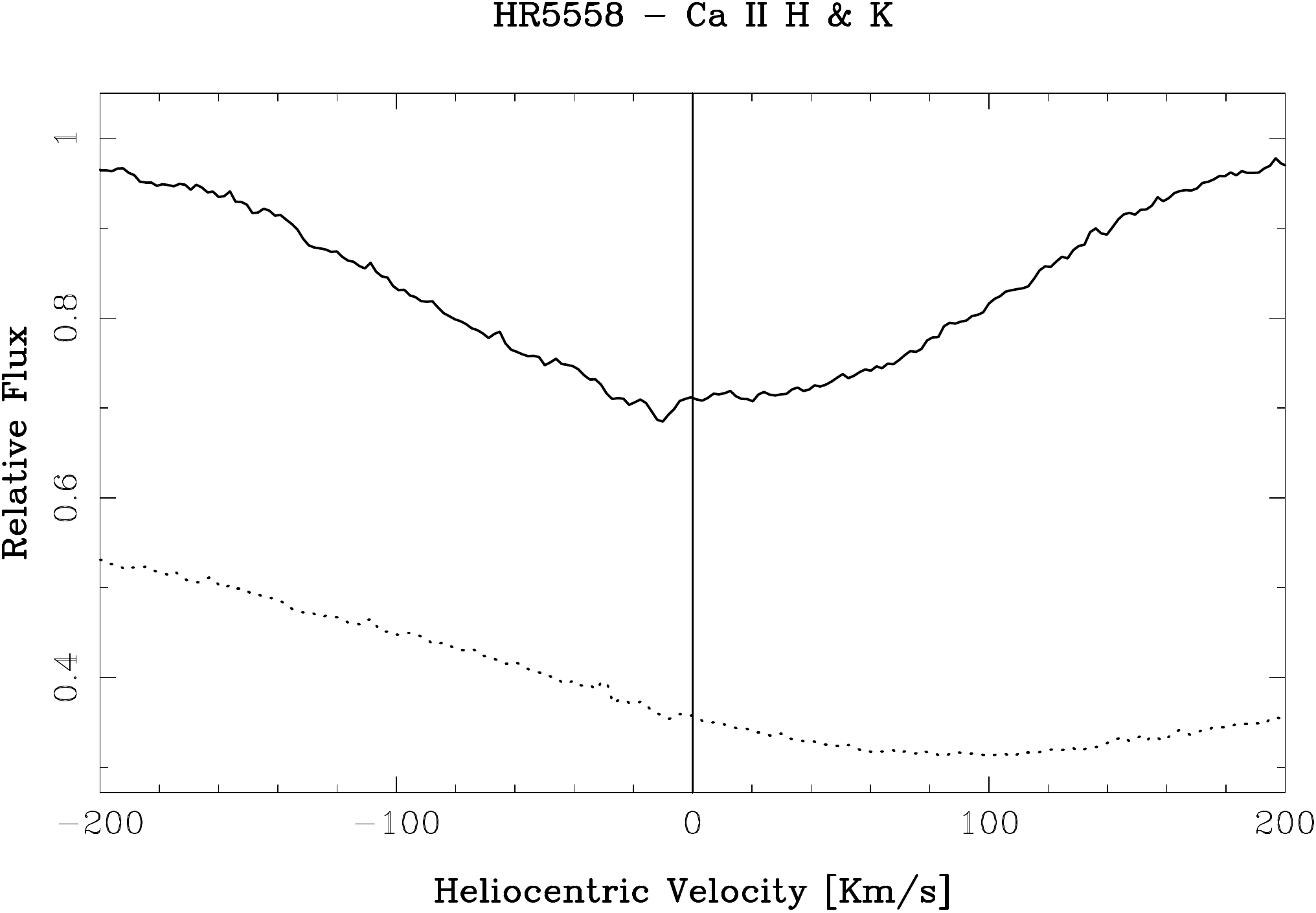}
\includegraphics[scale=0.3,angle=0,clip=true]{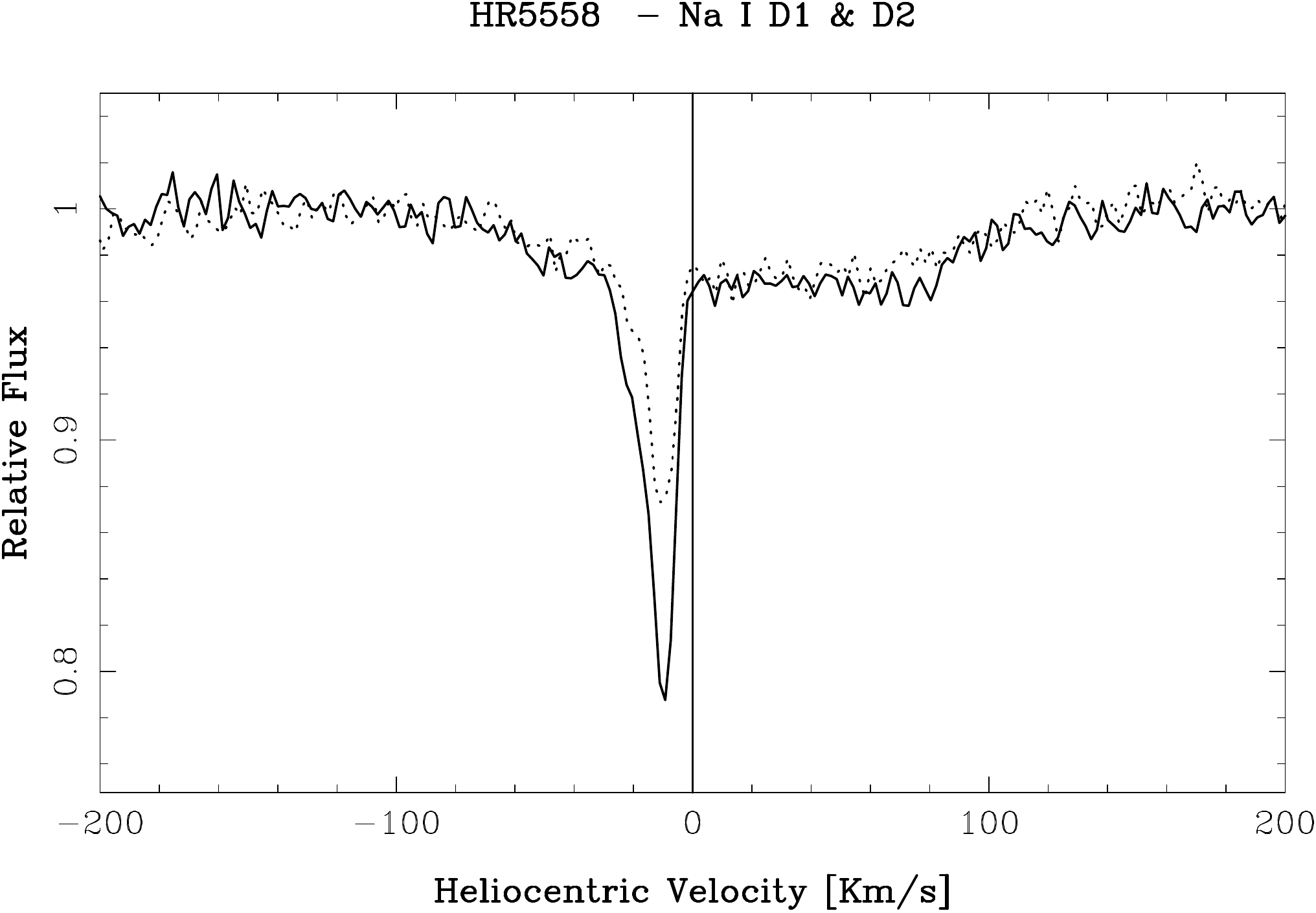}
\caption{Same as Figure~\ref{fig4} for spectral standard HR~5558}\label{fig4c}
\end{center}
\end{figure}

The fact that both the radial velocity of the star and of the
absorbing cloud are similar is not a sufficient condition to rule out
that the absorption lines detected are of interstellar origin. There
is not a unique criterion that may be used in order to discriminate
between an interstellar or circumstellar origin for absorption
features, instead it is common to refer to a set of criteria that can
be used together in order to resolve the nature of the absorption
features \citep{dun97b}. To rule out an interstellar origin for an
absorption feature one has to consider:

\begin{itemize}

\item Is the radial velocity of the narrow absorption feature the same
as the stellar velocity?

\item Is the radial velocity of the narrow absorption consistent or
not with known interstellar velocities in the same direction?

\item Is the observed $N$(Ca~{\sc ii})/$N$(Na~{\sc i}) column density
  ratio consistent with circumstellar (usually$>$1) or interstellar
  (usually$<$1) values?

\item Is the absorption observed towards stars in the similar line of sight ?
  

\end{itemize}
 

The $N$(Ca~{\sc ii})/$N$(Na~{\sc i}) must, however, be treated with
care as low density clouds or shocked shells in the interstellar
medium have also been found to exhibit $N$(Ca~{\sc ii})/$N$(Na~{\sc
 i}$>$1 \citep[e.g.][]{crawford91,smoker2015}.

When multiple epoch observations are available, one can also study
variability of the absorption features, as circumstellar lines often
vary with time \citep{beust98,redfield2007a,welsh2013}. However not
all circumstellar lines are variable, as for instance in $\beta$ Pic
the strongest component is stable and corresponds to the main disc
absorption \citep[as opposite to the variable components which are
  interpreted as evaporating comets;][]{beust98}.

We measured the equivalent widths and radial velocities of the narrow
absorption features on the normalized spectra by using the Emission
Line Fitting (ELF) Starlink routine, which fits Gaussian profiles to
the absorption features and returns the equivalent width and centre of
these unsaturated components. Table~\ref{ewnarrow} presents the
derived equivalent widths and radial velocities of the narrow Ca~{\sc
  ii} and Na~{\sc i} absorption features present in the spectra of our
targets.

\begin{table*}
\renewcommand{\arraystretch}{0.5}
\scriptsize
 \begin{center} 
\begin{tabular}{@{}c@{\qquad}cc@{\qquad}cc@{\qquad}cc@{\qquad}cc}
\\
\hline
\\
Star          & \multicolumn{2}{c}{Ca~{\sc ii}~K} & \multicolumn{2}{c}{Ca~{\sc ii}~H} &  \multicolumn{2}{c}{Na~{\sc i} D$_1$}& \multicolumn{2}{c}{ Na~{\sc i} D$_2$}  \\ \\
              & EW & V$_{\odot}$                  &  EW & V$_{\odot}$                 &   EW & V$_{\odot}$        &  EW & V$_{\odot}$                 \\
              & (m$\rm{\AA}$) & (km s$^{-1}$)                   &  (m$\rm{\AA}$) & (km s$^{-1}$)                  &   (m$\rm{\AA}$) & (km s$^{-1}$)         &  (m$\rm{\AA}$) & (km s$^{-1}$)                  \\
\hline	       					       										  
\hline\\

HD~61950      & 27.1$\pm 2.1$&  8.3$\pm 0.4$                          &   12.2$\pm 2.7$  &  8.8$\pm 0.7$      &    -    &  -   &   -   & -      \\     
              & 14.2$\pm 1.9$& 17.9$\pm 0.8$                          &    6.3$\pm 2.1$  & 17.1$\pm 0.9$     &    -    &  -   &   -   & -       \\
HD~75416      &  2.5$\pm 1.1$& -7.9$\pm 3.1$                          &   -              & -        &    -   &  -   &   -   & -       \\     
              &  2.7$\pm 0.5$&  6.9$\pm 0.5$                          &   -   & -        &    -   &  -   &   -   & -       \\     
              &  1.3$\pm 0.7$& 17.2$\pm 1.8$                          &   -   & -        &    -   &  -   &   -   &  -      \\     
HD~110058     & 13.9$\pm 2.1$& -1.2$\pm 0.5$                          &    5.8 $\pm 2.5$ & -2.4$\pm 0.6$     &   11.3$\pm 1.5$    & -1.4$\pm 0.9$ &  28.7$\pm 2.2$    &  -0.3$\pm 0.8$   \\  
              & 22.3$\pm 1.3$& 11.8$\pm 0.1$                          &   19.6$\pm 1.9$  & 11.9$\pm 0.3$     &   45.5$\pm 4.9$    & 12.9$\pm 0.7$ &  46.9$\pm 5.2$    &  12.0$\pm 0.5$   \\   
HD~166841     & 20.1$\pm 1.7$& 0.1 $\pm 0.2$                           &    6.1$\pm 2.1$  & 1.5 $\pm 0.4$     &   93.1$\pm 9.7$    &-0.1$\pm 0.3$ & 134.4$\pm 6.9$   &  -0.1$\pm 0.2$    \\     
HR~5588      
            
              & -            & -                                     &   -   &    -                         &    4.2$\pm 1.1$    &  -17.1$\pm 4.9$ &  13.1$\pm 0.9$   & -17.4$\pm 3.8$    \\     
              &  4.7$\pm 1.1$&-10.3$\pm 0.9$                        &   -   &    -                          &   20.5$\pm 3.2$    &  -10.5$\pm 0.8$ &  33.3$\pm 4.5$   & -10.2$\pm 0.6$    \\

\hline
\end{tabular}

\medskip 
\caption{Equivalent widths and heliocentric velocities of the narrow
absorption lines observed in the spectra of the target stars. The
equivalent widths are in m$\rm{\AA}$. For stars where two entries are
listed, these correspond to separate velocity components.}\label{ewnarrow}

\end{center}
\end{table*}
\normalsize

\subsection{Cloud modeling}

The Voigt Absorption Profile/Interstellar Dabbler software for
modeling interstellar absorption lines \citep[VAPID;][]{howarth2002},
was used to estimate the column density ($N$), radial velocity ($v$)
and the velocity dispersion ($b$) of the material causing the observed
absorption in the Ca~{\sc ii} K and Na~{\sc i} D$_2$ lines.  VAPID
assumes a Gaussian line-of-sight velocity distribution for each
absorbing cloud, and uses Voigt functions and least-squares
optimization to estimate cloud parameters of the many cloud components
required to reproduce the observed spectra. Uncertainties in the
resulting parameters are estimated using a Monte Carlo method to
derives confidence intervals in the parameter space. Oscillator
strengths of 0.635 and 0.631 were assumed for the Ca~{\sc ii} K and
Na~{\sc i} D$_2$ lines respectively \citep{morton1991}.

\begin{figure}
\begin{center}
\includegraphics[scale=0.3,angle=0,clip=true]{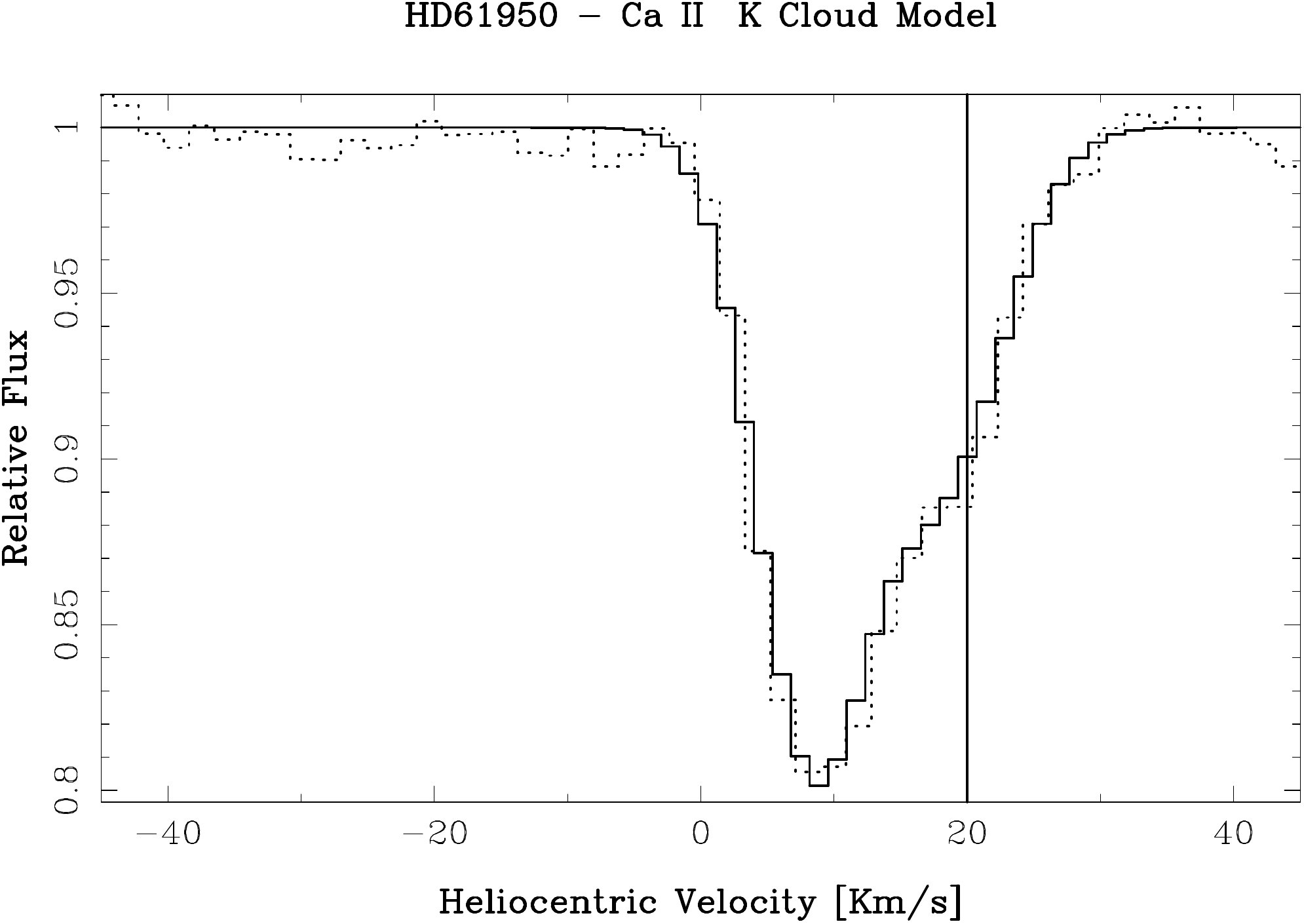}               
\includegraphics[scale=0.3,angle=0,clip=true]{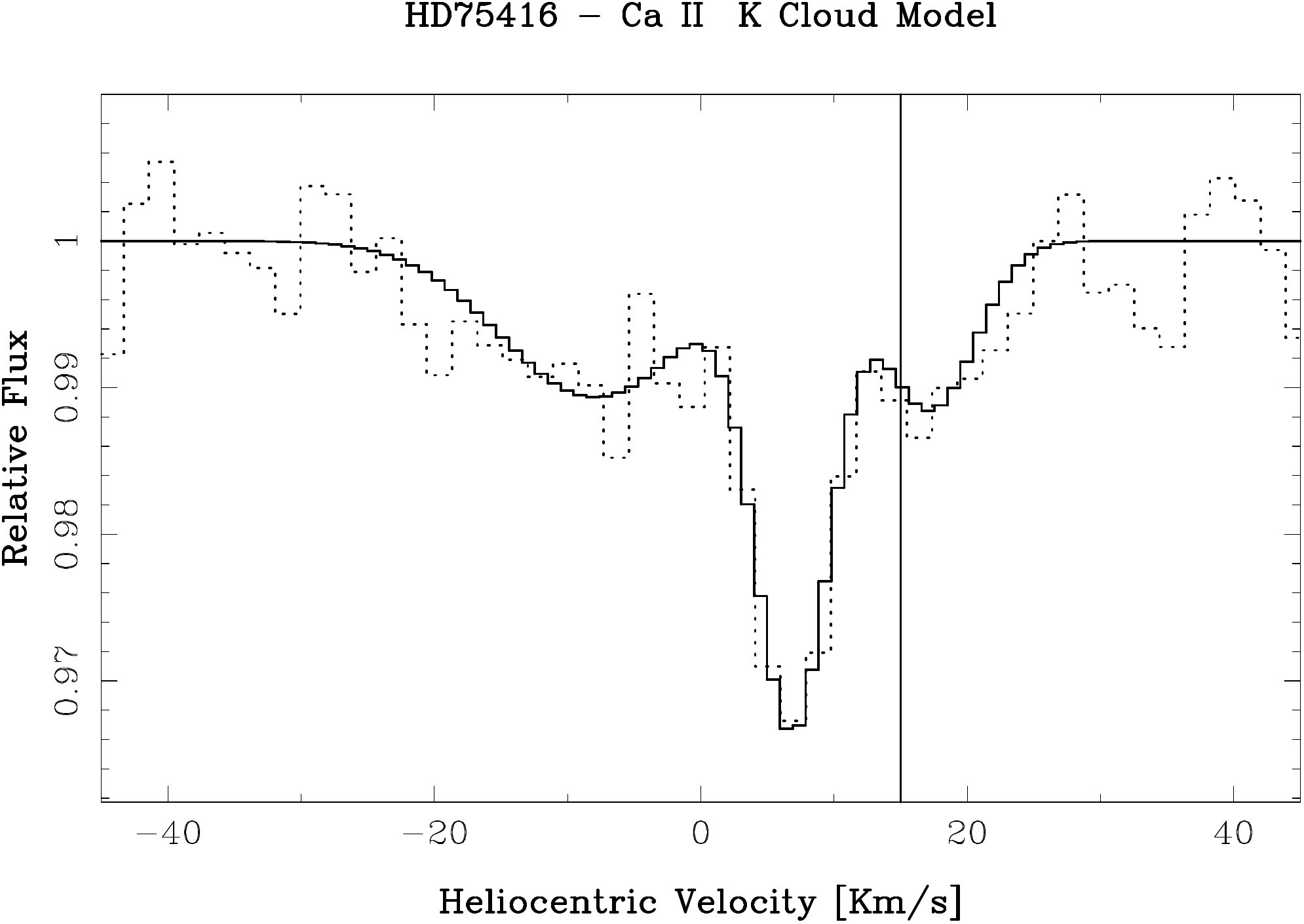}
\caption{Cloud models for the Ca~{\sc ii} K absorption features seen
  towards HD~61950 (top) and HD~75416 (bottom). The solid line shows
  the model that gives a satisfactory to the data (dotted line). The
  vertical line represents the heliocentric velocity of the
  star.}\label{hd75416model}
\end{center}
\end{figure}

\begin{figure*}
  \begin{center}

\includegraphics[scale=0.3,angle=0,clip=true]{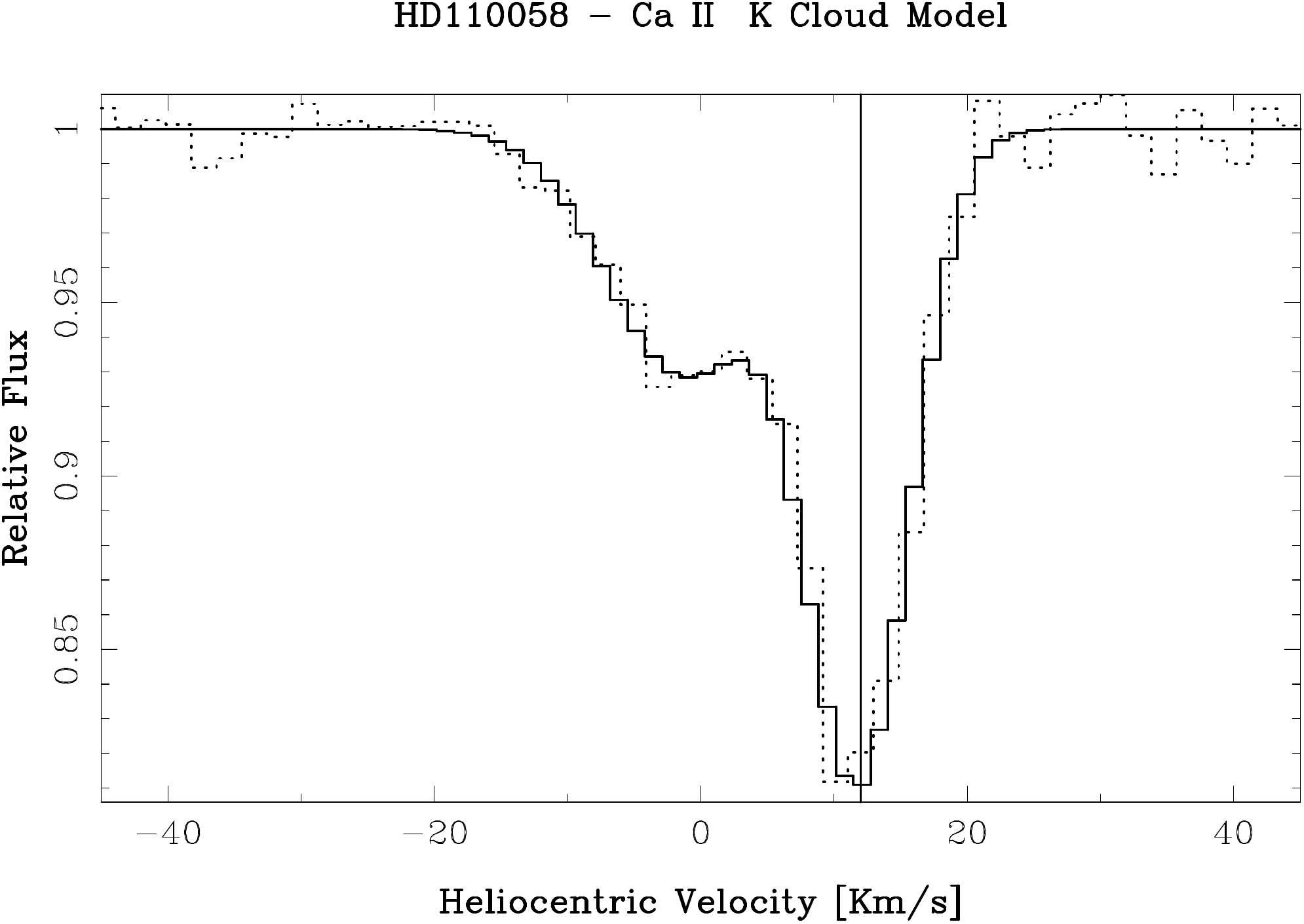}
\includegraphics[scale=0.3,angle=0,clip=true]{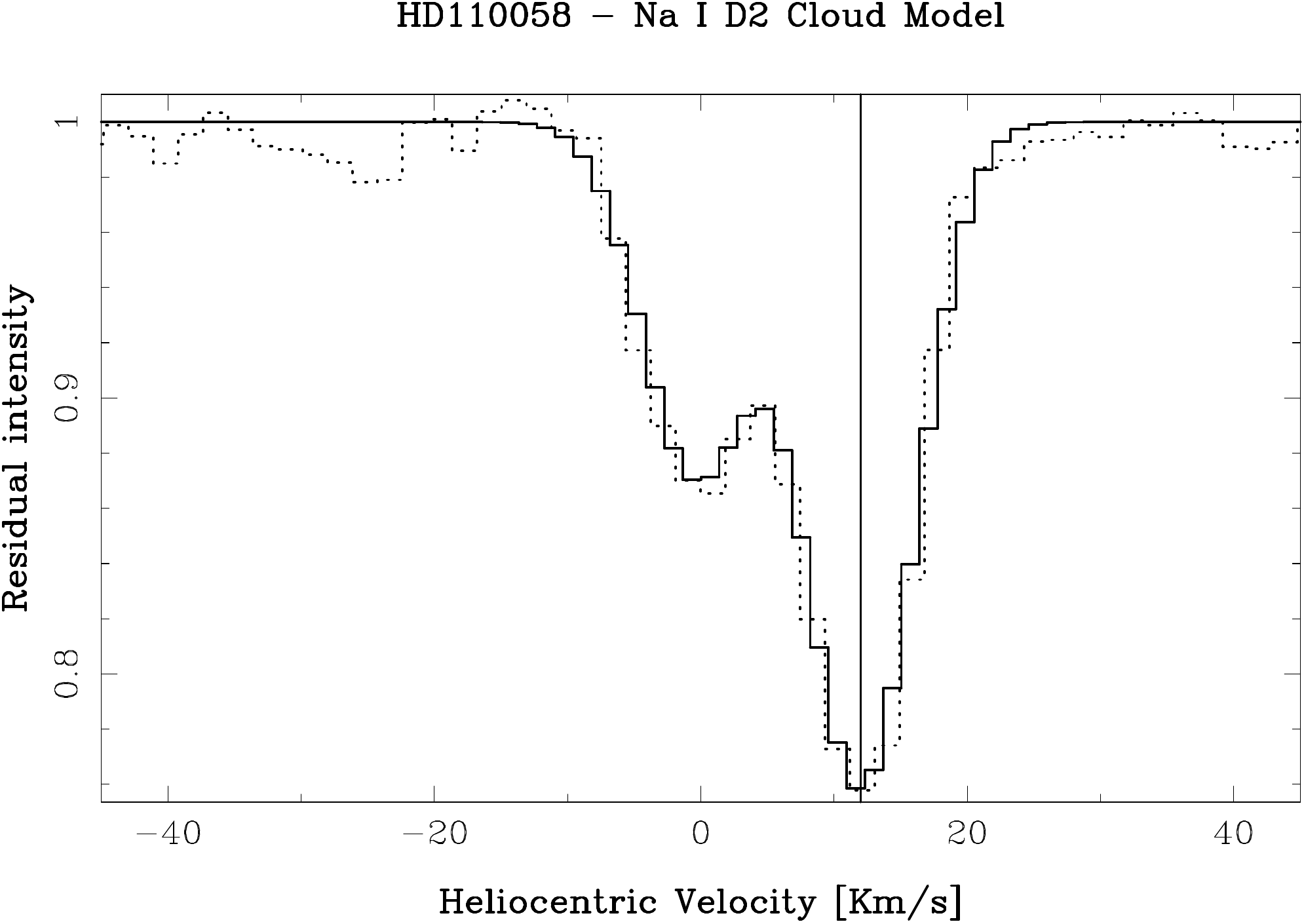}
\vspace{5mm}*      
\includegraphics[scale=0.3,angle=0,clip=true]{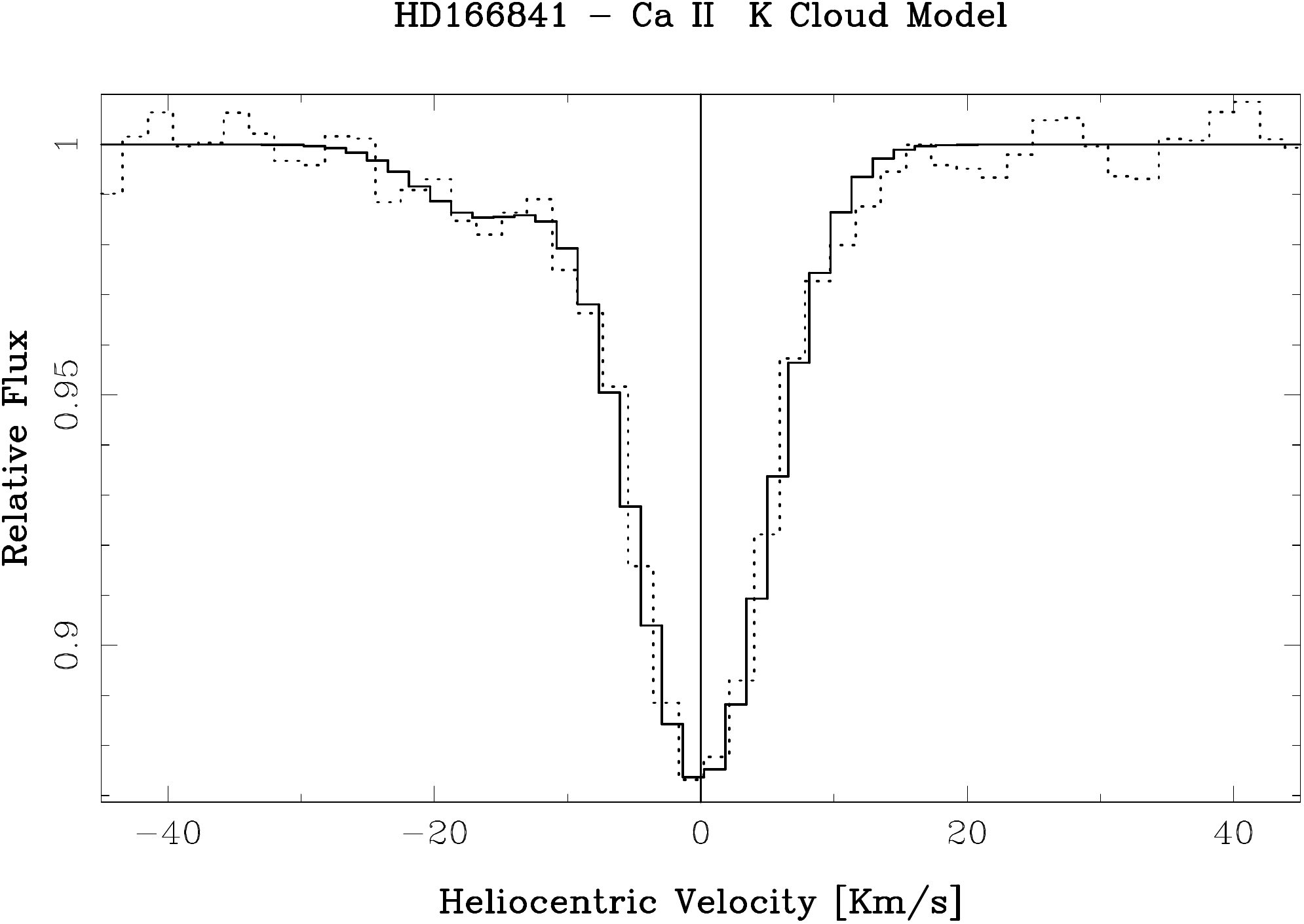}
\includegraphics[scale=0.3,angle=0,clip=true]{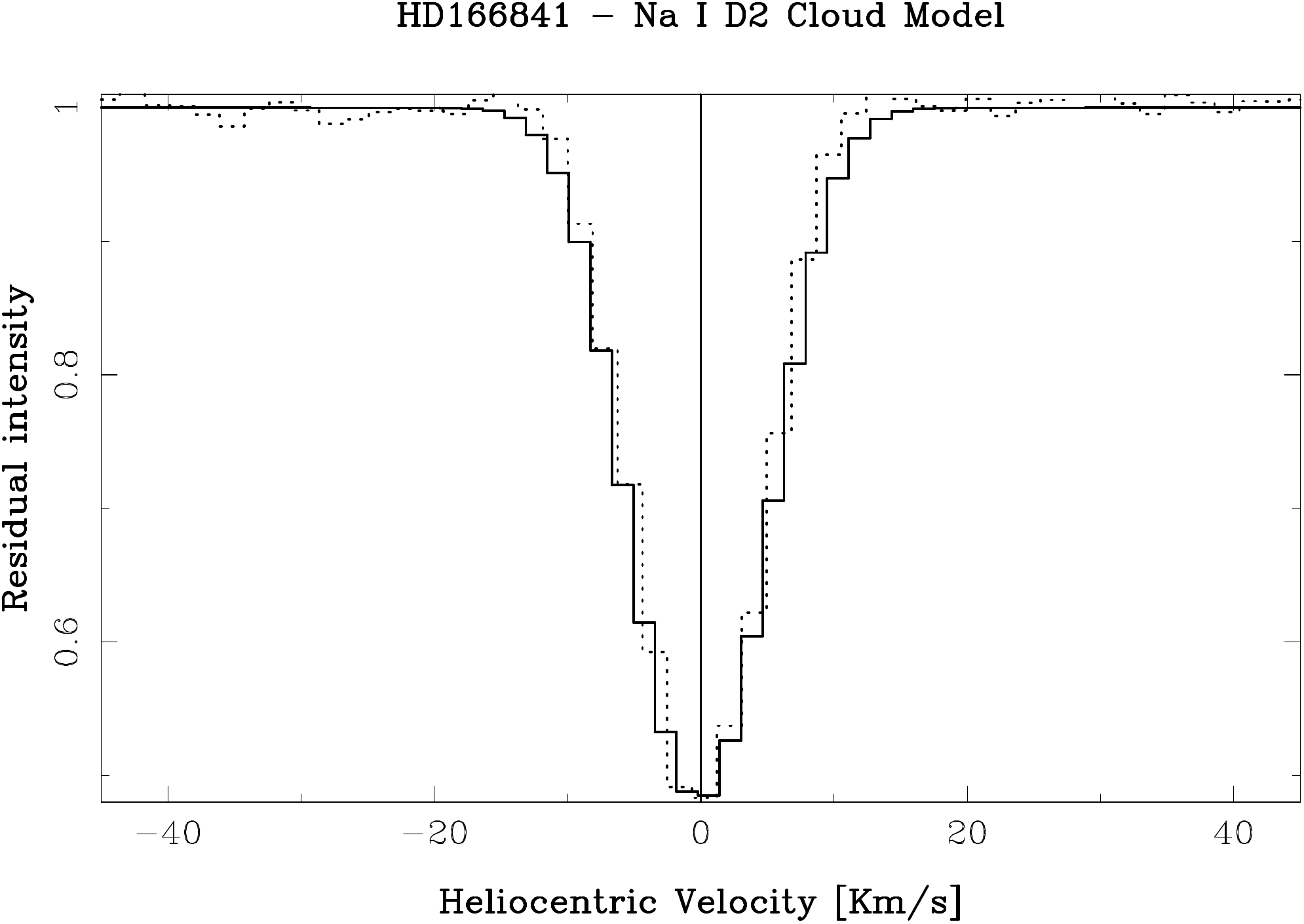}
\vspace{5mm}*  
\includegraphics[scale=0.3,angle=0,clip=true]{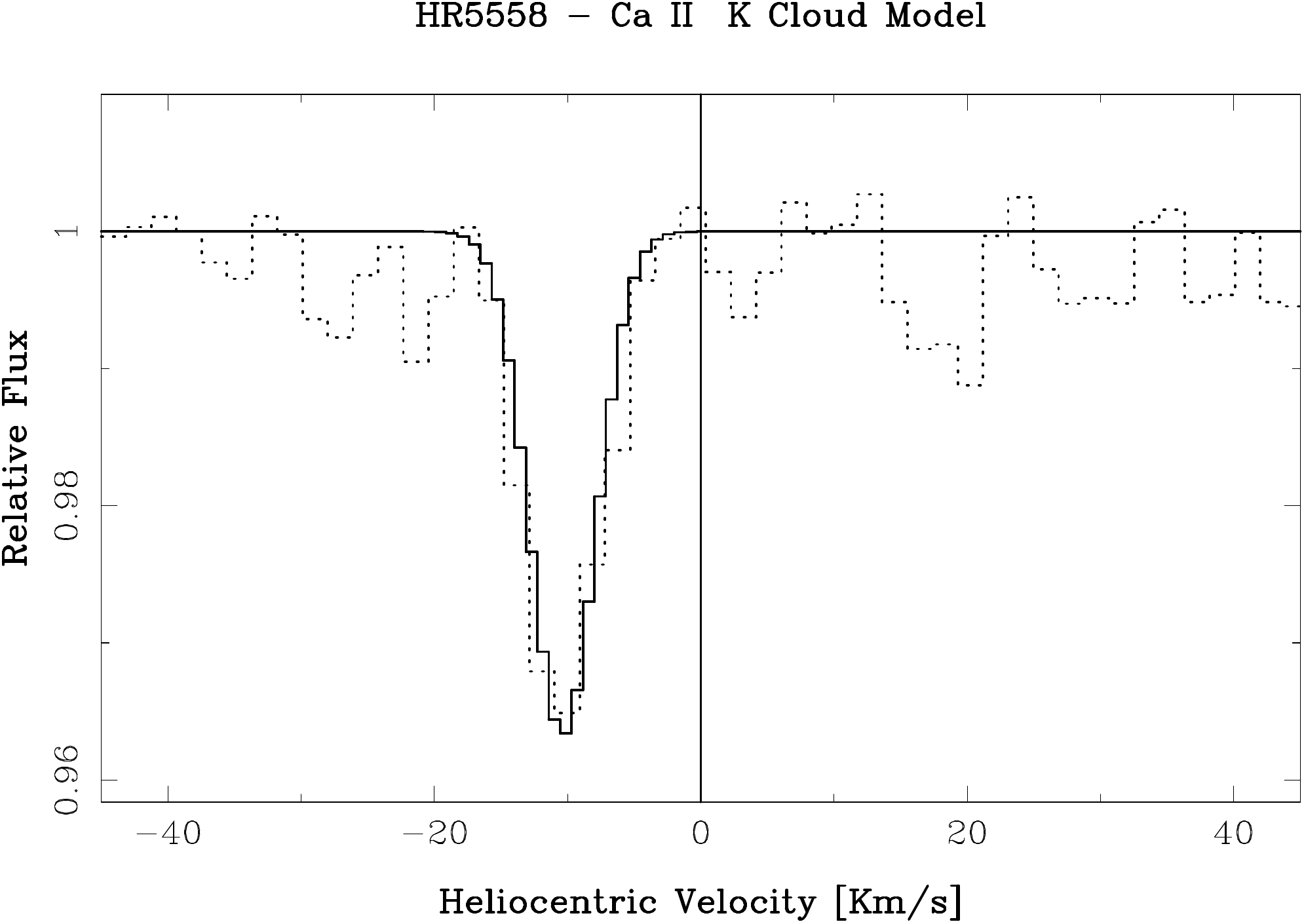}
\includegraphics[scale=0.3,angle=0,clip=true]{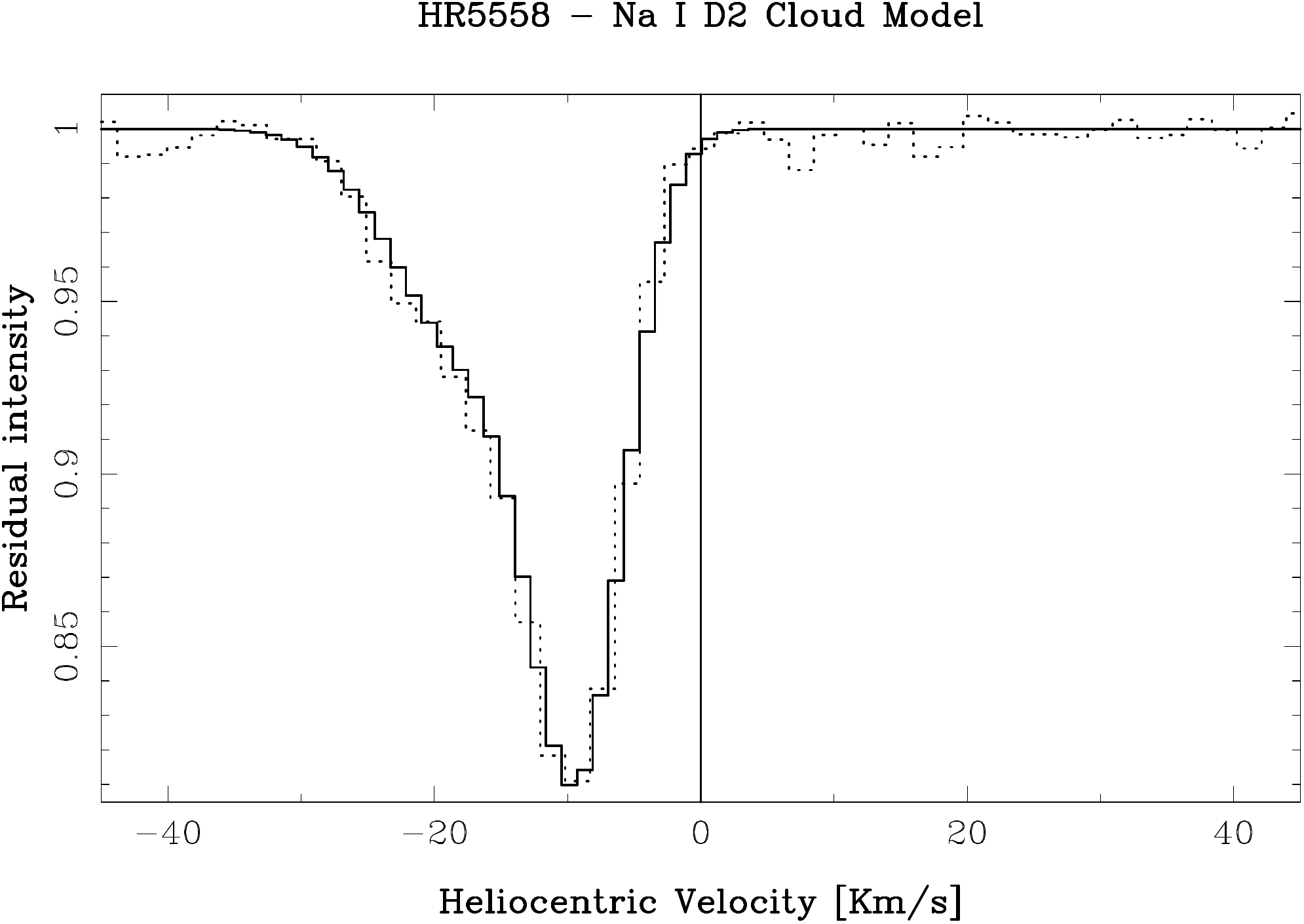}

\caption{Cloud models for the Ca~{\sc ii} K (left panel) and
 Na~{\sc i} D$_2$ (right panel) absorption features seen towards
HD~110058, HD~166841 and HR~5558. The solid line shows the model that
gives a satisfactory fit to the data (dotted line). The vertical line
represents the heliocentric velocity of the star.}\label{fig5}
\end{center}
\end{figure*}

The best-fit model parameters that were found to reproduce
successfully the data are presented in Table~\ref{clouds}, while
models versus data are plotted in Figures~\ref{hd75416model} and
\ref{fig5}.

%

\begin{table*}
\renewcommand{\arraystretch}{0.5}
\scriptsize
 \begin{center} 

\begin{tabular}{@{}c@{\qquad}ccc@{\qquad}ccc@{\qquad}c}
\\
\hline
\\
Star          & \multicolumn{3}{c}{Ca~K}         & \multicolumn{3}{c}{Na~$D_2$} & $N$(Ca~{\sc ii})/$N$(Na~{\sc i}) \\ \\
              & V$_{\odot}$ & b & $\log N  $                 &  V$_{\odot}$ & b & $\log N  $                \\              & (km s$^{-1}$) & (km s$^{-1}$)  & (cm$^{-2}$)                   & (km s$^{-1}$)  & (km s$^{-1}$)  & (cm$^{-2}$) \\
\hline	       					       				      				  
\hline\\

HD~61950     &8.3 $\pm 0.4$&1.9$\pm 0.6$ &11.56$\pm 0.03$ &-&-&-&-  \\     
             &18.1$\pm 0.7$&3.6$\pm 1.2$ &11.32$\pm 0.05$&-&-&-&-   \\
HD~75416   &-10.2 $\pm 1.8$&9.2$\pm 2.0$ &10.46$\pm 0.06$ &-&-&-&-   \\     
             &6.9 $\pm 0.4$&0.3$\pm 0.1$ &10.67$\pm 0.02$ &-&-&-&-   \\     
             &18.5$\pm 1.4$&0.2$\pm 0.5$ &10.20$\pm 0.04$ &-&-&-&-   \\     
HD~110058   & -1.4$\pm 0.4$&5.7$\pm 0.9$ &11.19$\pm 0.03$& -0.1$\pm 0.2$& 2.1$\pm 0.7$ &11.19$\pm 0.02$ &1.00  \\  
             &11.8$\pm 0.2$&0.8$\pm 0.1$ &11.73$\pm 0.03$& 12.2$\pm 0.1$& 2.0$\pm 0.4$ &11.53$\pm 0.02$ &1.41  \\   
HD~166841    & 0.1$\pm 0.1$&4.4$\pm 0.4$& 11.40$\pm 0.02$& -0.25$\pm 0.1$& 3.5$\pm 0.2$& 12.01$\pm 0.02$ &0.25   \\
            &-16.5$\pm 1.9$&3.9 $\pm 3.4$& 10.43$\pm 0.11$&- &- &- &-                                         \\     
HR~5588     &- &- &-                                    &-19.1$\pm 0.5$& 3.9$\pm 0.9$& 10.87$\pm 0.04$ &-    \\     
            &-10.4$\pm 0.4$&0.4$\pm 0.3$& 10.70$\pm 0.02$&-9.6$\pm 0.2$& 0.6$\pm 0.1$ &11.69$\pm 0.02$ &0.1  \\


\hline
\end{tabular}

 \caption{Derived line profile parameters for narrow Ca~K and Na~
   D$_2$ velocity components derived from VAPID modeling. The
   parameter errors reported represent the 1$\sigma$ dispersion in
   each parameter (i.e., single-parameter 68\% confidence
   intervals). The last column gives the $N$(Ca~{\sc ii})/$N$(Na~{\sc
     i}) ratio.  }\label{clouds}


\end{center}
\end{table*}
\normalsize

%
%

%
%
%

\clearpage

\ifpdf

\begin{figure}
\begin{center}
\includegraphics[scale=0.7,angle=0]
{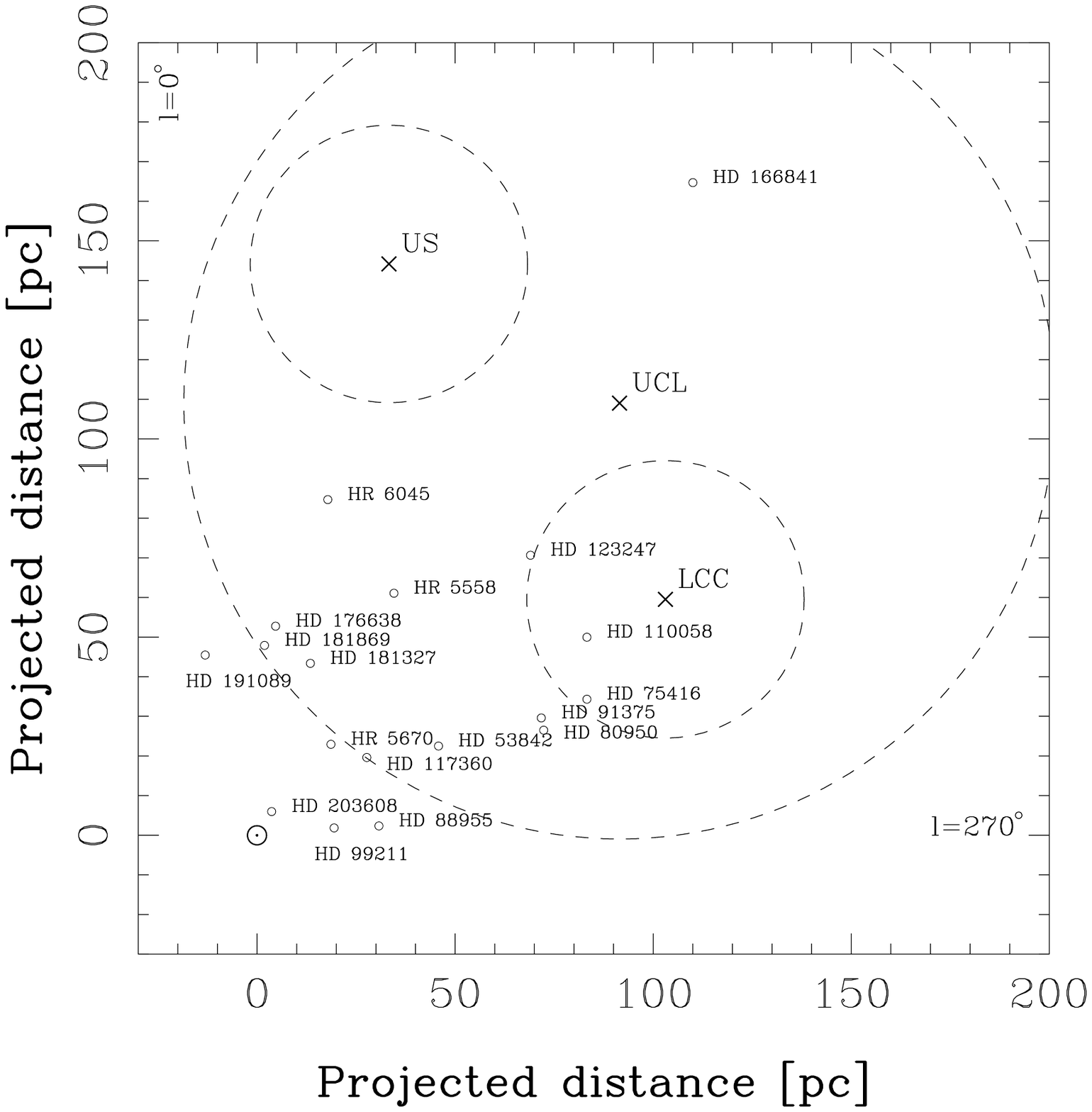}
\caption{Projection of the positions of the target stars (open
circles) onto the Fourth Quadrant of the Galactic plane. The dashed
circles indicate the location of the UCL, LCC and US H~{\sc i} shells
as described in \citet{degeus89}, \citet{degeus92} and
\citet{dezeeuw99}. The centre of each shell is indicated with a
cross. The position of the Sun is located at the origin of the
coordinates, and is marked with the $\odot$ symbol. The star HD~61950
is located too far away to appear in the plot, positioned near
$(x=330,y=63)$.}\label{4th}
\end{center}
\end{figure}

\fi

\clearpage

\subsection{Velocity Projections of known IS clouds/shells}

All the stars in this study lie in the Fourth Galactic Quadrant. This
region is almost entirely occupied by the Scorpius-Centaurus
association, extending between $\sim 290^{\rm o}-360^{\rm o}$ in
longitude and being the closest OB association to the Sun
\citep{blaauw1964,degeus89}. Three subgroups have been identified to
reside within this association: the Upper Centaurus-Lupus (UCL), Lower
Centaurus-Crux (LCC) and the Upper Scorpius (US) shells
\citep{degeus92} . These subgroups have been studied extensively, and
the interstellar medium in their proximity has been modeled as
expanding spherical shells of gas, whose physical parameters such as
shell centre, radius and expansion velocity have been derived by
\citet{degeus89}, \citet{degeus92} and \citet{dezeeuw99}. The location
of our target stars with respect to the UCL, LCC and US shells are
plotted in Figure~\ref{4th}.\\

In addition to the UCL, LCC and US shells, one must also consider the
structure of the local interstellar medium including the Local
Interstellar Cloud \citep[LIC;][]{redfield2015}. The LIC is a warm,
low density interstellar cloud which is itself located within a hot
Local Cavity in the interstellar medium which extends to approximately
50~pc \citep{welsh2010}. The LIC is thought to move in the direction
$l=186^{\rm o}\pm3$, $b=-16^{\rm o}\pm3$, with a heliocentric velocity
of $v=26\pm 1$ km~s$^{-1}$ \citep{lallement95}.  The Local Cavity is
thought to contain several small clouds located within a few tens of
parsecs of the Sun, with characteristics similar to the LIC. One such
possible cloud is the `G' cloud, which appears to move at $v=29\pm 1$
km~s$^{-1}$ in the direction $l=184.5^{\rm o}\pm2.3$ and $b=-20.5^{\rm
  o}\pm1.8$ \citep{lallement92}.


\subsubsection{The LIC and G clouds}

Given an interstellar cloud moving with velocity $v_w$ in a direction
$(l_w,b_w)$, the line of sight velocity component of the cloud in a
direction $(l,b)$ is given by \citep{crutcher82}:
\begin{equation}\label{lic}
 \frac{v}{v_w}=\sin{b}\cdot\sin{b_w}+\cos{b}\cdot\cos{b_w}\cdot\cos{(l-l_w)},
\end{equation}
where the terms on the right-hand side correspond to the cosine
of the angle between $(l,b)$ and $(l_w,b_w)$ .

\subsubsection{The UCL, LCC and US shells}

For an expanding shell model the following equation can be used to
determine the velocity component for a point $(l,b)$ on the surface of
a sphere of radius $r_s$, centred at a distance R from the Sun at
$(l_o,b_o)$ \cite[e.g.][]{crawford91}:
\begin{equation}\label{shells}
 v=\pm\frac{v_o}{r_{s}}\sqrt{R^{2}(\cos{\theta}^{2}-1)+r{_s}^2},
\end{equation}
where $\cos{\theta}$ is given by the term in the right-hand side of
Equation~\ref{lic} (substituting $(l_w,b_w)$ for $(l_o,b_o)$),
$\theta$ is the angle between $(l,b)$ and $(l_o,b_o)$ and $v_o$ is the
expansion velocity with respect to the Local Standard of Rest (LSR) or
V$_{\rm LSR}$. Conversion from LSR to Heliocentric velocities was
performed using the RV Starlink package. We used equations \ref{lic}
and \ref{shells} to calculate the projected velocity components for
the different interstellar clouds and shells in the direction of our
targets with detected absorption, which are presented in
Table~\ref{isclouds}.


\begin{table*}
\setlength{\tabcolsep}{5mm}
\renewcommand{\arraystretch}{0.5}
\scriptsize
 \begin{center} 
\begin{tabular}{@{}ccccc}
\\
\hline
\\
Star         & LIC   &  G     & UCL        & LCC \\
             & (km~s$^{-1}$) &(km~s$^{-1}$)& (km~s$^{-1}$) &(km~s$^{-1}$)     \\
 
\hline	      							       										  
\hline\\      						       											  
 
HD61950      &    0.65            & 0.91          &  9.44 &   -\\
HD75416      &    -3.91           & -4.01         &  4.28 &   -\\
HD110058     &    -11.94          &-14.16         & -5.25 &  1.92\\
HD166841     &    -15.00          &-15.81         & -9.06 &  -\\
HR5558       &    -21.55          &-24.70         & -4.93  &  \\

\hline
\end{tabular}
\caption{Heliocentric Velocity Projections of known interstellar clouds/shells in the direction of the stars presenting narrow absorption features.}
\label{isclouds}
 
\medskip


\end{center}
\end{table*}
\normalsize

\subsection{Discussion of individual stars}

\subsubsection{HD~61950}

HD~61950 is a B8 dwarf that shows a distinctive two-component
absorption feature seen in both the Ca H and K lines.  The absorption
components are centred at +8.3 and +18.1 km~s$^{-1}$, lying near the
base of the broad photospheric line. The red-most component, at
+18.1$\pm 0.7$ km~s$^{-1}$ coincides within the errors with the star's
heliocentric velocity (20$\pm 5$ km~s$^{-1}$), although there is quite
a large uncertainty on the latter due to the lack of prominent
metallic lines on the spectrum of this B-type star.  The Ca~{\sc ii}
equivalent width ratio suggests that for both narrow components the
line is unsaturated, as the derived K:H EW ratio is $\sim 2$ in both
cases (a Ca~{\sc ii} K:H ratio of 2 is expected in the case of
unsaturated lines, as the oscillator strength of the H-line is half
that of the K-line). \\

Given that HD~61950 is the most distant star in our sample
(d$=362$~pc), it is more probable that these features are of
interstellar origin.  The velocity projections of known clouds and
shells predicts that the UCL shell should contribute an absorption
feature at +9.4~km~s$^{-1}$ (Table~\ref{isclouds}), that coincides
extremely well with the main absorption component seen at
+8.3~km~s$^{-1}$.

Comparison with Ca~{\sc ii} K observations toward stars in similar
directions suggests that both components in the spectrum of HD~61950
are interstellar. The LSR velocities of the two narrow Ca~{\sc ii}
absorption components seen toward HD~61950 (l=281$^{\rm o}$,
b=-21$^{\rm o}$, d=362~pc) correspond to -3.7~km~s$^{-1}$ and
+5.9~km~s$^{-1}$. \citet{hunter06} detected two Ca~{\sc ii} K
absorption components toward HD~76131 (l=273$^{\rm o}$, b=-7$^{\rm
  o}$, d=453~pc) at LSR velocities of -2~km~s$^{-1}$ and
+8~km~s$^{-1}$, with EWs of 110~m$\rm\AA$ and 21~m$\rm\AA$
respectively. These two velocity components are also detected in the
spectra of HD~67536 (l=276$^{\rm o}$, b=-16$^{\rm o}$, d=450~pc), at
LSR velocities of -4~km~s$^{-1}$ and +5.3~km~s$^{-1}$ with EW of
16~m$\rm\AA$ and 11~m$\rm\AA$ respectively. The LSR velocities of the
two components detected toward HD~61950 coincide with the range of
velocities of the two interstellar clouds reported by \citet{hunter06}
are detected. It is therefore very likely that both absorption
features seen toward HD~61950 are interstellar in nature.

%
%
%

\subsubsection{HD~75416}

HD~75416 ($\eta$~Cha) lies at a distance of 97~pc and is also a B8V
star. It is the brightest member of the 8~Myr old $\eta$~Cha cluster
\citep{mamajek1999}, and is the only debris disk in this young, disk
rich, stellar association \citep{sicilia2009}. Gaussian modeling of
the photospheric lines indicate that the star is a fast rotator
($v\,\sin{i}\sim$290~km~s$^{-1}$), as was already noted by
\cite{sicilia2009}, and that the stellar heliocentric velocity is +15
$\pm5$~km~s$^{-1}$ (in agreement with the
V$_{\odot}=14\pm10$~km~s$^{-1}$ value derived by
\citet{debruijne2012}). A very weak absorption feature is detected at
the centre of the photospheric Ca~{\sc ii} K profile. Three separate
components at heliocentric velocities of -10.2, +6.9 and
+18.5~km~s$^{-1}$, were required in order to satisfactorily fit the
observed profile (Figure~\ref{hd75416model}). There is no evidence of
corresponding features in the intrinsically weaker Ca~{\sc ii} H
line.\\

The cloud models imply very low calcium column densities, with the
strongest components being the ones at -10.2 and +6.9~km~s$^{-1}$. The
velocity component observed at +6.9~km~s$^{-1}$ could be related to
the UCL shell (velocity projection of +4.3~km~s$^{-1}$), however no
known clouds/shells can account for the 18.5~km~s$^{-1}$
component. This feature is interesting because, despite being very
weak (only 1.3~m$\rm\AA$ equivalent width), it lies close to the
stellar velocity. However, the lack of additional information
regarding this absorption feature makes it difficult to draw any
conclusions regarding the nature of this feature. Additional
observations of the Na~{\sc i} D region could help in resolving the
origin of these features


\subsubsection{HD~110058}

HD~110058 is an A0V star located at a distance of $107_{-8}^{+10}$~pc,
in the direction of the LCC shell \citep{Leeuwen2007}.  HD~110058 was
first identified by \citet{mann98} as a debris disc host star.  The
disc was undetected at 1350~$\mu$m with SCUBA by \citet{syl01}, which
allowed them to estimate an upper limit to the dust mass of less than
$5\times10^{-6}\,M_{\odot}$. \citet{syl01} estimate the fractional
disc luminosity of $1.89\times10^{-3}$, in  good agreement with
our estimate of $2.0\times10^{-3}$. This is very similar to the dust
fractional luminosity of $\beta$~Pic
\cite[$2.4\times10^{-3}$,][]{moor2011}, and comparable to those of
other A-type star debris disc around which circumstellar gas has been
detected
\citep{Hughes08,moor2011,Dent14,moor2015,Lieman2016,marino2016}.

Near-infrared Very Large Telescope/SPHERE imaging has revealed that,
similarly to $\beta$~Pic, the debris disc around HD~110058 is seen
very close to edge-on \citep{kasper2015}. Recent ALMA observations
confirm the near to edge-on inclination of the disc, and also show a
5-$\sigma$ detection of carbon monoxide towards the system
\citep{Lieman2016}.

There is some discrepancy in the literature regarding HD~110058's
radial velocity. \citet{moor2006} measured the radial velocity of
HD~110058 to be +21.7$\pm1.3$~km~s$^{-1}$, while \citet{debruijne2012}
estimate a radial velocity of 5$\pm1$~km~s$^{-1}$. Our measurement
lies somewhere in between the estimates of \citet{moor2006} and
\citet{debruijne2012}.



A strong, sharp, absorption component is detected in both the Calcium
and Sodium regions at the very base of the photospheric line. Modeling
of the absorption feature indicates that there are actually two
velocity components, centred near -1~km~s$^{-1}$ and +12~km~s$^{-1}$
respectively. None of the known interstellar clouds/shells appear to
have projections at these velocities, and so cannot account for the
observed features. The  component near +12~km~s$^{-1}$, lies
right at the stellar radial velocity of +12$\pm3$~km~s$^{-1}$ we
derived, which would argue for a circumstellar nature for this
feature.\\

It is interesting to note that in the case of the +12~km~s$^{-1}$
component, the equivalent widths of the D$_1$ and D$_2$ lines are
equal within the errors, suggesting saturation. The same effect is
seen with regards to the H and K calcium lines. This could be
explained by the presence of clumpy intervening material, similar to
what has been modeled for $\beta$ Pictoris  \citep{lagrange92}. For both velocity components, the
Ca~{\sc ii}/Na~{\sc i} ratios that we derive are close to unity, and
are the largest found in our sample. \citet{redfield2007a} notes
however that is is hard to differentiate a circumstellar from
interstellar origin based on the abundance ratios alone, since a wide
range of Ca~{\sc ii}/Na~{\sc i} ratios are detected even locally. In
their search for atomic absorption in nearby debris discs,
\citet{redfield2007a} find that their best candidates for
circumstellar absorption have Ca~{\sc ii}/Na~{\sc i} ratios between
3.9 and 46.\\

The coincidence between the radial velocities of the narrow absorption
feature and the star, the lack of known interstellar clouds/shells
that can account for absorptions at these velocities, together with
the Ca~{\sc ii}/Na~{\sc i} ratio of $\sim1$ strongly argue for a
circumstellar nature for the component seen at +12~km~s$^{-1}$. The
fact that the component seen at $\sim$12~km~s$^{-1}$ is saturated
while the one at $\sim$-1~km~s$^{-1}$ is not, indicates that they have
different column densities, suggesting that they may be different in
nature.

The MIKE observations of HD~110058 and of reference stars in the
vecinity of HD~110058 confirm that the feature at a lower velocity is
interstellar. The feature is observed in all stars
(Figure~\ref{figMike}), and its depth increases with distance as
expected if the absorption is caused by interstellar material. On the
other hand, the absorption feature at $\sim$12~km~s$^{-1}$ is detected
only towards HD~110058. Therefore we conclude that this absorption
feature is indeed circumstellar.

%

\ifpdf

\begin{figure*}
  \begin{center}

\includegraphics[scale=0.4]{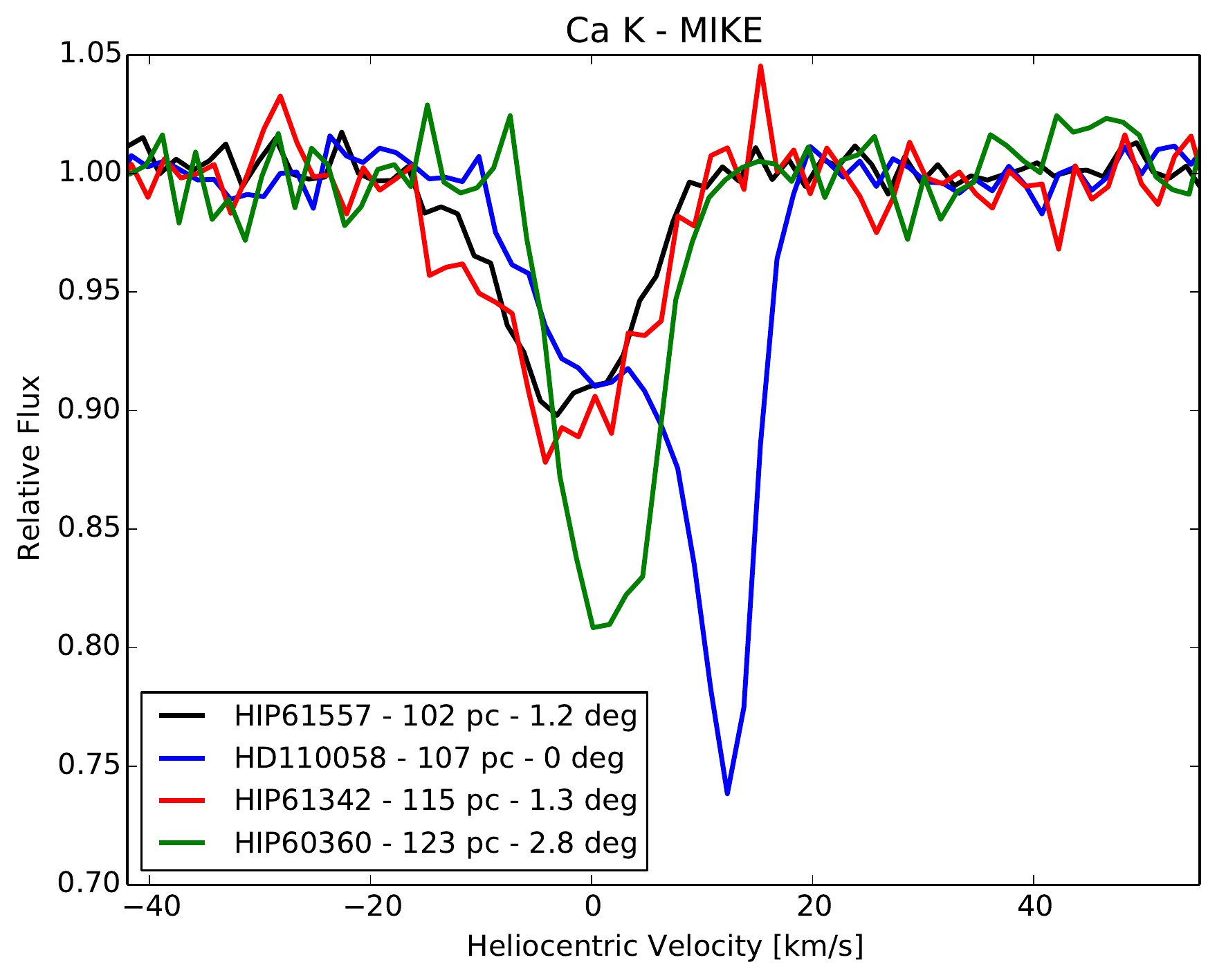}\hspace*{-.4em}
\includegraphics[scale=0.4]{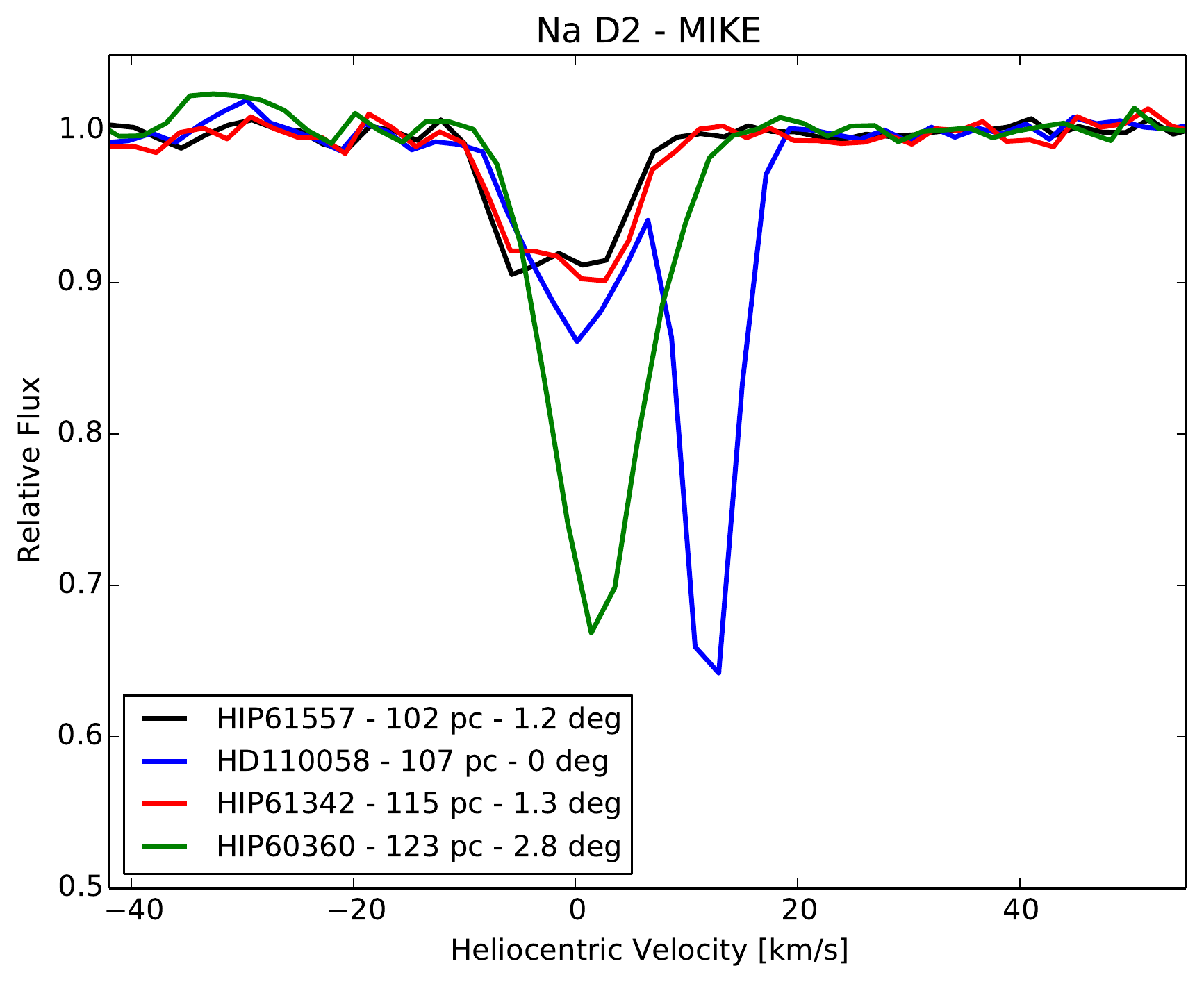}\hspace*{-2.2em}

\caption{MIKE Ca~{\sc ii} spectra towards HD~110058 and reference
  targets for interstellar absorption characterization. The
  parallactic distance to the reference stars and their angular
  separation from HD~110058 are annotated in the
  figure.}\label{figMike}
\end{center}
\end{figure*} 

\fi

\subsubsection{HD~166841}

HD~166841 is B9V star located 214~pc away from the Sun in the
direction of the centre of the UCL shell. The star was first
identified as a debris disc host  by \citet{mann98}, but no further
studies of this object have been carried out. The value of the
projected rotational velocity that we derive (245 $\pm8$~km~s$^{-1}$)
is consistent with the star being of late B-type. Despite being a
quite distant star, the E(B-V) value we estimate assuming the normal
colours of a B9V star is quite low (E(B-V)=0.02).\\

A strong narrow absorption feature is easily seen right at the centre
of the photospheric Ca~{\sc ii} and Na~{\sc i} lines. Both the calcium
and sodium absorption features are found to be located at
V$_{\odot}$=0$\pm1$~km~s$^{-1}$, coinciding extremely well with the
radial velocity of the star. The derived equivalent widths of the
different component suggest that the lines are not saturated and cloud
modeling indicates that Ca~{\sc ii}/Na~{\sc i} column density ratio is
$\sim 0.3$.\\

Although the velocity projections of known clouds/shells cannot be
responsible for the observed absorption feature, we investigated the
presence of similar absorption components in the spectra of stars in
similar directions. The LSR velocity of the component seen toward
HD~166841 (l=326$^{\rm o}$, b=-22$^{\rm o}$, d=214~pc) is
-1.8~km~s$^{-1}$. \citet{hunter06} reports +2.4$\pm0.1$~km~s$^{-1}$
Ca~{\sc ii} K component on the direction of HD~142758 (l=325$^{\rm
o}$, b=-4$^{\rm o}$, d=4000~pc) with 65~m$\rm\AA$ equivalent width.
\citet{hunter06} also detected two Ca~{\sc ii} K absorption features
in the spectrum of HD~143448 (l=324$^{\rm o}$, b=6$^{\rm o}$,
d=520~pc), at -3.6$\pm0.1$~km s$^{-1}$ (EW=18~m$\rm\AA$) and
+2.0$\pm0.6$~km~s$^{-1}$ (EW=59~m$\rm\AA$), both referred to the
LSR. These two stars are located at a higher Galactic latitude than
HD~166841 and are more distant, but it is worth noting that the first
component seen toward HD~143448 is similar within the errors to the
one seen toward HD~166841, both in equivalent width and radial
velocity.\\

Despite being located right at the centre of the photospheric lines,
there is not enough supporting evidence to attribute this feature to a
circumstellar origin. The star is quite distant and there is some evidence of a
correlated velocity structure between the feature seen toward
HD~166841 and those towards stars in nearby lines of sight. One
argument that could favor a circumstellar nature for the feature seen toward
HD~166841 is the fact that when interstellar absorption components are detected
toward field stars, generally more than one component is detected,
whereas in the case of HD~166841 only one distinctive narrow
absorption is observed. Observations of stars with lines of sight
closer to HD~166841 than the stars quoted above could help to
clarify the nature of HD~166841's narrow absorption feature.

\subsubsection{HR~5558}\label{hr5558}

HR~5558 is an A0V spectral standard located 76~pc away from the Sun.
We serendipitously detected a narrow absorption feature noticeable in
the Na~{\sc i} D lines (EW of 20~m$\rm\AA$ and 33~m$\rm\AA$ in the D$_1$
and D$_2$ lines respectively), and less evident in the Ca~{\sc ii} K
line. The feature was not detected in the Ca~{\sc ii} H line.\\

There is no correspondence of the radial velocity of the absorption
feature with the radial velocity of the star, suggesting that the
absorption is of interstellar origin. This is supported by the low Ca~{\sc ii}/Na~{\sc i} ratio
derived. Given the diagnostics above, plus the fact that
HR~5558 is not known to be a debris disc host star, we conclude that the absorption
feature seen toward this star is interstellar in origin.

\section{Conclusions}

We have conducted a search for optical circumstellar absorption lines
in the spectra of 16 debris disc host stars. We found no evidence of
emission line activity, confirming their more evolved and quiescent
evolutionary state. Four stars show narrow absorption features close
to the centre the photospheric Ca~{\sc ii} and Na~{\sc i} D lines. In
addition similar absorption features were detected in the spectrum of
one our spectral standards.

Of the four stars showing narrow absorption features, two are younger
than 17~Myr (the other two stars do not have available ages in the
literature). This is consistent with the findings of \cite{welsh2013} in
which they note that the stars that exhibit circumstellar gas activity
are significantly younger than quiescent systems.


We also find that the systems towards which absorption is detected are
all fast rotators. This has been discussed as a possible proxy for
detecting gas absorption in debris disks, since it can be indicative
of a system seen close to edge-on \citep{welsh2013}. With the
exception of HD~176638 none of the stars with
$v\,\sin{i}$<140~km~s$^{-1}$ show signatures of gas absorption. We
also note that gas absorption is only detected toward A- or late
B-type stars, in agreement with previous detections of exocomet
bearing systems \citep[e.g.][]{redfield07b,welsh2013,eiroa2016}.

The features detected toward HD~61950 and HR~5558 can be almost
unambiguously attributed to interstellar absorption, while in the case
of HD~75416 and HD~166841 the association is less clear. The lack of
known interstellar absorption features within the velocity ranges of
the lines seen in both HD~75416 and HD~166841 (which coincide with the
stellar velocities) leaves open the question of the nature of these
features. HD~75416 ($\eta$~Cha) is known to host an evolved
protoplanetary disk, possibly transitioning into the debris disk phase
\citep{sicilia2009}.  Re-observation of these two sources in order to
search for variability, together with observations of stars in
adjacent lines of sight, could help to disentangle the origin of these
features. The non-detection of atomic gas absorption towards the
CO-rich HD~181237 debris disc system is consistent with this system
being seen close to pole-on \citep{marino2016}.

The most compelling evidence for a circumstellar gas is seen in one of
the two velocity components of the absorption seen in the spectrum of
HD~110058 (at +12~km~s$^{-1}$). The good agreement with the stellar
velocity, the Ca~{\sc ii}/Na~{\sc i} column density ratio close to
unity, the lack of known interstellar clouds or shells at the correct
velocities in HD~110058's direction, and the fact that the disc is
observed close to edge-on strongly suggest that the absorption is
caused by atomic gas present in HD~110058's disc. The additional MIKE
data provides compelling evidence that the component at
+12~km~s$^{-1}$ arises in the circumstellar environment of
HD~110058. 

The detection of optical absorption gas features towards HD~110058,
adds to the many similiarities with the $\beta$ Pic system (age, dust
fractional luminosity, orientation, and presence of both atomic and
molecurlar gas).  In the latter, the gas is believed to be of
secondary origin, i.e. produced by the release of atomic and molecular
species by volatile-rich bodies \citep{Dent14,kral2016}. Future
studies of this new $\beta$ Pic-like system could provide further
information on the origins of gas in debris discs.

\section*{Acknowledgements}

A.S.H. carried out part of this work while being funded by the PPARC
GeminiFundacion Andes UK/Chile studentship program. A.S.H. thanks
Rafael Brahm for useful discussions on calibration of MIKE data. The
National Radio Astronomy Observatory is a facility of the National
Science Foundation operated under cooperative agreement by Associated
Universities, Inc. SC acknowledges support from Millennium Science
Initiative, Chilean Ministry of Economy: Nucleus P10-022-F.








%
%
%

\bsp	
\label{lastpage}
\end{document}